\documentclass[pra,twocolumn,superscriptaddress,showpacs,amsmath,amssymb,citeautoscript, floatfix]{revtex4-1}

\usepackage[T1]{fontenc}
\usepackage[utf8]{inputenc}
\usepackage{lipsum}
\usepackage{amsmath}
\usepackage{amssymb}
\usepackage{bbm}
\usepackage{braket}
\usepackage{xcolor}
\usepackage{pifont}
\usepackage[mathscr]{euscript}
\usepackage[shortlabels]{enumitem}
\usepackage[justification=justified]{subcaption}
\usepackage{graphicx}
\usepackage{lipsum}
\allowdisplaybreaks
\usepackage{float}
\usepackage{graphicx}
\usepackage{dsfont}
\usepackage{comment}
\usepackage[colorlinks=true]{hyperref}  
\hypersetup{
    bookmarks=true,         
    unicode=false,          
    pdftoolbar=true,        
    pdfmenubar=true,        
    pdffitwindow=false,     
    pdfstartview={FitH},    
    pdftitle={Machine Learning Nematic Order},    
    pdfauthor={Sobral et al.},     
    pdfsubject={},   
    pdfcreator={},   
    pdfproducer={}, 
    pdfkeywords={} {} {}, 
    pdfnewwindow=true,      
    colorlinks=true,       
    linkcolor=blue, 
    citecolor=blue,        
    filecolor=magenta,      
    urlcolor=blue           
}

\newcommand{\equref}[1]{Eq.~(\ref{#1})}
\newcommand{\equsref}[2]{Eqs.~(\ref{#1}) and (\ref{#2})}
\newcommand{\secref}[1]{Sec.~\ref{#1}}
\newcommand{\figref}[1]{Fig.~\ref{#1}}
\newcommand{\refcite}[1]{Ref.~\onlinecite{#1}}

\newcommand{\appref}[1]{Appendix~\ref{#1}}

\usepackage{bm}
\newcommand{\pdagger}{{\phantom{\dagger}}}

\renewcommand{\approx}{\simeq}

\renewcommand{\vec}[1]{\boldsymbol{#1}}

\definecolor{wrongultramarine}{rgb}{1,0.5,0}

\begin{document}

\title{Machine Learning Microscopic Form of Nematic Order \texorpdfstring{\\}{Lg} in twisted double-bilayer graphene}

\author{João Augusto Sobral}
\affiliation{Institute for Theoretical Physics, University of Innsbruck, Innsbruck A-6020, Austria}

\author{Stefan Obernauer}
\affiliation{Institute for Theoretical Physics, University of Innsbruck, Innsbruck A-6020, Austria}

\author{Simon Turkel}
\affiliation{Department of Physics, Columbia University, New York, New York 10027, USA}

\author{Abhay N. Pasupathy}
\affiliation{Department of Physics, Columbia University, New York, New York 10027, USA}
\affiliation{Condensed Matter Physics and Materials Science Division, Brookhaven National Laboratory, Upton, New York 11973, USA}

\author{Mathias S.~Scheurer}
\affiliation{Institute for Theoretical Physics, University of Innsbruck, Innsbruck A-6020, Austria}

\begin{abstract}
Modern scanning probe techniques, like scanning tunneling microscopy (STM), provide access to a large amount of data encoding the underlying physics of quantum matter. In this work, we analyze how convolutional neural networks (CNN) can be employed to learn effective theoretical models from STM data on correlated moiré superlattices. These engineered systems are particularly well suited for this task as their enhanced lattice constant provides unprecedented access to intra-unit-cell physics and their tunability allows for high-dimensional data sets within a single sample. Using electronic nematic order in twisted double-bilayer graphene (TDBG) as an example, we show that including correlations between the local density of states (LDOS) at different energies allows CNNs not only to learn the microscopic nematic order parameter, but also to distinguish it from heterostrain.  These results demonstrate that neural networks constitute a powerful methodology for investigating the microscopic details of correlated phenomena in moiré systems and beyond.

\end{abstract}

\maketitle

\section{Introduction}
Driven by the impressive improvements in machine learning (ML) in the last couple of years, exploring its potential for quantum many-body physics has recently become subject of intense research \cite{carleoMachineLearningPhysical2019, dawidModernApplicationsMachine2022}. For instance, ML provides powerful tools to solve inverse problems that occur frequently in physics \cite{leeMachinelearningSpectralFunction2023,
duboisUntrainedPhysicallyInformed2022a,
berthusenLearningCrystalField2021,
chertkovComputationalInverseMethod2018}:
given a model, it is often straightforward with conventional many-body techniques to compute observables that can be measured experimentally, whereas the often needed inverse problem of extracting the model and underlying microscopic physics from observations is much more challenging and typically even formally ill-defined.  
A second example of a large class of applications of ML in physics is ML-assisted analysis of experiments, in particular of those yielding image-like data like scanning tunneling microscopy (STM) \cite{choudharyComputationalScanningTunneling2021, 
jouckenDenoisingScanningTunneling2022b,
wangMachineLearningIdentification2020, zhangMachineLearningElectronicquantummatter2019}, photoemission \cite{liuRemovingGridStructure2022}, and others \cite{khanUsingCycleGANsGenerate2023, 
chenMachineLearningOptical2022, edeDeepLearningElectron2021, 
iwasawaUnsupervisedClusteringIdentifying2022,
burzawaClassifyingSurfaceProbe2019, basakDeepLearningHamiltonians2022}.

In the context of applying ML algorithms to data from imaging techniques like STM, van der Waals moir\'e superlattices \cite{balentsSuperconductivityStrongCorrelations2020,andreiGrapheneBilayersTwist2020} are particularly promising for three reasons: (i) they display a huge variety of correlated quantum-many-body phenomena, such as interaction-induced insulating phases \cite{caoCorrelatedInsulatorBehaviour2018}, magnetism \cite{sharpeEmergentFerromagnetismThreequarters2019}, superconductivity \cite{caoUnconventionalSuperconductivityMagicangle2018}, electronic nematic order \cite{kerelskyMaximizedElectronInteractions2019,jiangChargeOrderBroken2019,choiElectronicCorrelationsTwisted2019,caoNematicityCompetingOrders2021}, which can also coexist microscopically \cite{caoNematicityCompetingOrders2021,linZerofieldSuperconductingDiode2022}. Despite intense research on these phenomena over several decades, e.g., in the pnictides or cuprates, their origin and relations are still subject of ongoing debates. However, compared to these microscopic crystalline quantum materials, moir\'e superlattices are (ii) highly tunable; for instance, the density of carriers can be varied within a single sample just by applying a gate voltage (as opposed to chemical doping) and even the interactions can be tuned \cite{liuTuningElectronCorrelation2021}. This allows producing large data sets of measurements on a single sample, containing a lot of information on the microscopic physics. This aspect, which is crucial for data-driven approaches, is further enhanced by (iii) the large moir\'e unit cells of these systems compared to that of microscopic crystals, increasing the relative spatial resolution of scanning-probe techniques significantly. This enables experiments to probe the structure of the wave functions within the unit cell, and thus provides unprecedented access to the microscopic physics compared to conventional quantum materials. For instance, in the extreme limit of only one degree of freedom (Wannier state or pixel) per unit cell, the broken rotational symmetry of the electron liquid---the defining property of electronic nematic order \cite{fradkinNematicFermiFluids2010,fernandesWhatDrivesNematic2014}---is not visible as a consequence of translational symmetry and thus requires a careful analysis of the behavior around impurities \cite{goetzDetectingNematicOrder2020}. 

\begin{figure}[tb]
	\centering{}\includegraphics[width=\columnwidth]{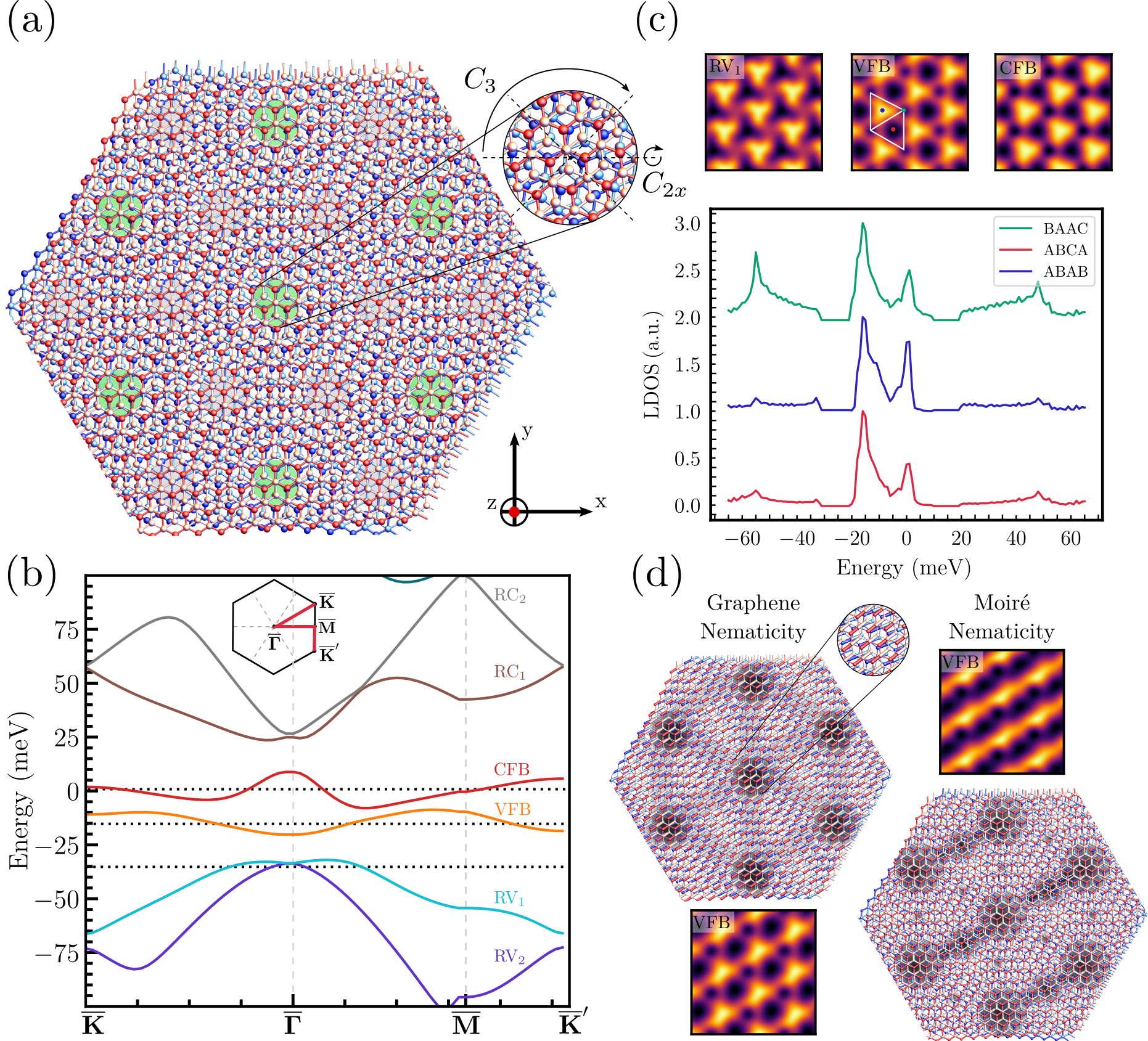}\caption{ (a) Representation in real space of the TDBG heterostructure. Green highlighted domains emphasize the emerging moir\'e pattern due to the combination of two AB-stacks of graphene bilayers with a relative twist angle $\theta$. (b) Band structure at $\theta=1.05^{\circ}$ along highly symmetrical points from the moir\'e Brillouin zone (inset). Solid lines represent conduction and valence flat bands (CFB/VFB) as well as remote bands (R) with valley $\eta=+$. The chemical potential corresponds to roughly a half-filling fraction ($\nu=0.475$) of the CFB.  (c) LDOS for three fixed energies [black dotted lines in (b)] as a function of position (top), and for varying energy at fixed high-symmetry positions (bottom) in the moir\'e unit cell (white rhombus).  (d) Schematic real-space illustration of two limiting cases of graphene and moir\'e nematicity, along with two sample LDOS plots at a fixed energy in the VFB; both show clear $C_{3}$ breaking.
 }
	\label{fig:bandmoire}
\end{figure}

In this work, we explore these advantages of moir\'e superlattices for extracting or `learning' effective field-theoretical descriptions of their correlated many-body physics from STM data. This can be viewed as an inverse problem and is also conceptually related to the goal of `Hamiltonian learning' in quantum simulation \cite{granadeRobustOnlineHamiltonian2012,wiebeHamiltonianLearningCertification2014,wangExperimentalQuantumHamiltonian2017,  valentiHamiltonianLearningQuantum2019, kokailQuantumVariationalLearning2021,yuPracticalEfficientHamiltonian2022}, albeit in rather different regimes and based on different measurement schemes. As a concrete example, we use electronic nematic order in twisted double-bilayer graphene (TDBG) \cite{caoTunableCorrelatedStates2020, liuTunableSpinpolarizedCorrelated2020, shenCorrelatedStatesTwisted2020,rubio-verduMoireNematicPhase2022,heChiralitydependentTopologicalStates2021,kuiriSpontaneousTimereversalSymmetry2022,suSuperconductivityTwistedDouble2022}. This  moir\'e system consists of two AB-stacked bilayers of graphene that are twisted against each other; as one can see in \figref{fig:bandmoire}(a), it exhibits the point group $D_3$, generated by three-fold rotation $C_3$ along the out-of-plane $z$ axis and two-fold rotation $C_{2x}$ along the in-plane $x$ axis. Evidence of electronic nematic order has been observed in previous STM experiments \cite{rubio-verduMoireNematicPhase2022,samajdarElectricfieldtunableElectronicNematic2021} which clearly exhibit stripe-like features breaking the $C_3$ symmetry spontaneously for certain electron concentrations. While simple limiting cases have been compared with the data in \refcite{samajdarElectricfieldtunableElectronicNematic2021}, there is no systematic analysis of the microscopic form of nematicity in the system. To fill this gap, we consider the more general case in which all leading terms on the graphene and moir\'e scale describing nematic order in a continuum-model description of TDBG are included. In addition, as it is common in graphene moir\'e systems \cite{huderElectronicSpectrumTwisted2018,kerelskyMaximizedElectronInteractions2019,jiangChargeOrderBroken2019,choiElectronicCorrelationsTwisted2019,rubio-verduMoireNematicPhase2022}, we also allow for finite strain. The Hamiltonian defining the changes in TDBG resulting from nematic order and strain depends on a set of parameters $\beta$, which we reconstruct from STM data using convolutional neural networks (CNN) in a supervised learning procedure. As such, our study differs significantly from recent works, which focused on detecting the presence or absence of nematic order \cite{goetzDetectingNematicOrder2020} or performed a phenomenological data analysis of STM measurements \cite{tarantoUnsupervisedLearningTwocomponent2022} with ML, rather than extracting the underlying microscopic physics as we do here. 

\section{Results}
\subsection{Nematic order in TDBG}
\label{sec:nematicorderintdbg}
The non-interacting band structure of TDBG features two moir\'e minibands per spin and valley close to charge neutrality, where a variety of correlation-driven phenomena can emerge \cite{caoTunableCorrelatedStates2020, liuTunableSpinpolarizedCorrelated2020, shenCorrelatedStatesTwisted2020,rubio-verduMoireNematicPhase2022,heChiralitydependentTopologicalStates2021,kuiriSpontaneousTimereversalSymmetry2022,suSuperconductivityTwistedDouble2022}. In  \figref{fig:bandmoire}(b), these minibands are denoted as valence (VFB) and conduction flat bands (CFB). The band structure shown is obtained from continuum model calculations close to half filling of the CFB (band filling $\nu=0.475$), where electronic nematic order was observed to be the strongest \cite{rubio-verduMoireNematicPhase2022}, see \appref{app:DetailsModel} for more details. STM experiments probe the band structure and wave functions of a system by providing direct access to the spatial and energy dependence of the LDOS. Most commonly, the LDOS is studied either for a fixed position $\vec{r}_0$ over a range of different energies, $\mathcal{D}_{\vec{r}_0}(\omega)$, or for a fixed energy $\omega_0$ covering a spatial region of the system, $\mathcal{D}_{\omega_0}(\mathbf{r})$. The behavior of $\mathcal{D}_{\omega_0}(\mathbf{r})$ and $\mathcal{D}_{\vec{r}_0}(\omega)$ following from the continuum model for TDBG for three different energies and high-symmetry positions in the moir\'e unit cell is shown in \figref{fig:bandmoire}(c). The $C_3$ rotational and translational symmetry of the moir\'e lattice can be clearly seen in $\mathcal{D}_{\omega_0}(\mathbf{r})$. Meanwhile, $C_{2x}$ is broken, albeit weakly, as a consequence of the electric field required to control the electron filling to be close to the middle of the CFB in an open-faced STM sample geometry \cite{rubio-verduMoireNematicPhase2022}.

In graphene moir\'e systems, there are two fundamentally distinct sources of $C_3$ symmetry breaking---strain and electronic nematic order. Postponing the discussion of the former to below, electronic nematic order \cite{fradkinNematicFermiFluids2010,fernandesWhatDrivesNematic2014} refers to the spontaneous rotational symmetry breaking as a result of electronic correlations. While recent works also indicate the possibility of nematic charge-density wave states in TDBG  \cite{wilhelmNoncoplanarMagnetismTopological2022,heChiralitydependentTopologicalStates2021}, where moir\'e translational symmetry is simultaneously broken, we here focus on translationally symmetric nematic order since the STM data of \refcite{rubio-verduMoireNematicPhase2022} preserves moir\'e translations. The underlying nematic order parameter we study is a time-reversal- and moir\'e-translation-invariant vector $\mathbf{\Phi}=\Phi \hat{\mathbf{\Phi}}_\varphi$, $\hat{\mathbf{\Phi}}_\varphi=\left(\cos 2\varphi, \sin 2\varphi\right)$, transforming under the irreducible representation $E$ of $D_3$ (or of $C_3$, taking into account the weak $C_{2x}$ breaking); $\Phi$ and $\varphi$ stand for the intensity and orientation of the nematic director, respectively.
The microscopic form of nematicity can be modeled by a coupling of $\mathbf{\Phi}$ to a fermionic bilinear and reads in its most general form in a continuum-model description as \cite{samajdarElectricfieldtunableElectronicNematic2021}
\begin{equation}
\begin{gathered} 
\label{eq:nematorder1}
\mathcal{H}_{\bm{\Phi}}=\int_{\bm{r}}\int_{\Delta\bm{r}}\bm{\Phi}\cdot\bm{\phi}_{\sigma, \ell,s, \eta;\:\sigma^{\prime}, \ell^{\prime},s^{\prime}, \eta^{\prime}}\left(\bm{r},\Delta\bm{r}\right)\,\\ \times c_{\sigma,\ell,s,\eta}^{\dagger}\left(\bm{r}+\Delta \bm{r}\right)c_{\sigma^{\prime},\ell^{\prime},s^{\prime},\eta^{\prime}}^{\:}(\bm{r})+\text{H.c.},
\end{gathered}
\end{equation}
where $c^{\dagger}$  and $c$ are the electronic creation and annihilation operators.
This general form encompasses couplings between the two sublattices $s=A,B$ of the microscopic graphene sheets, the four graphene layers $\ell=1,\dots, 4$, the valley $\eta = \pm$ and spin $\sigma =\: \uparrow, \downarrow$ degrees of freedom in the tensorial form factor $\bm{\phi}_{\sigma,\ell,s, \eta;\:\sigma^{\prime},\ell^{\prime},s^{\prime}, \eta^{\prime}}(\bm{r},\Delta\mathbf{r})$; its two components are required to transform in the same way as $\vec{\Phi}$ under all symmetries of the system. In the following, we will take $\vec{\phi}$ to be trivial in the spin and diagonal in the valley indices, $\bm{\phi}_{\sigma,\ell,s, \eta;\:\sigma^{\prime},\ell^{\prime},s^{\prime}, \eta^{\prime}}=\delta_{\sigma,\sigma'}\delta_{\eta,\eta'}\bm{\phi}_{\ell,s;\:\ell^{\prime},s^{\prime}}(\eta)$. This is motivated by the weak spin-orbit coupling in graphene \cite{kaneQuantumSpinHall2005, minIntrinsicRashbaSpinorbit2006} and the lack of indications of interaction-induced spin-orbit coupling, which is also strongly constrained \cite{kiselevLimitsDynamicallyGenerated2017}. Furthermore, the intervalley-coherent nematicity is known to lead to stronger effects on the remote bands \cite{samajdarElectricfieldtunableElectronicNematic2021} that were not observed experimentally \cite{rubio-verduMoireNematicPhase2022}. 

Since we are working with a continuum theory, the space of possible couplings $\vec{\phi}$ in \equref{eq:nematorder1} is technically infinite dimensional. As such, a complete reconstruction of $\vec{\phi}$ from experimental data is impossible given the finite resolution and energy range of the available data. On top of this, it is not required either as we are primarily interested in understanding the low-energy behavior of the system. In the spirit of gradient expansions commonly used in continuum low-energy field theories, we will therefore only keep the leading terms in $\vec{\Phi}$. There is, however, a subtlety associated with the presence of an additional moir\'e length scale. We will therefore have to consider two basic classes of nematic orders, referred to as graphene (GN) and moiré (MN) nematicity \cite{rubio-verduMoireNematicPhase2022, samajdarElectricfieldtunableElectronicNematic2021}. 

In the case of MN, nematic order is associated with the moir\'e scale, i.e., we choose $\Delta \vec{r} = \mathbf{R}_{m_{1},m_{2}}=m_{1}\mathbf{L}_{1}^{M}+m_{2}\mathbf{L}_{2}^{M}$ in \equref{eq:nematorder1}, $m_{j}\in \mathbb{Z}$, with moir\'e lattice vectors $\mathbf{L}_{j}^{M}$, to represent the non-trivial transformation behavior of $\vec{\phi}$ under $C_3$. We can thus take it to be diagonal in the remaining internal indices, yielding 
\begin{equation}
\begin{gathered}
\label{eq:nematorder2}
    \mathcal{H}_{\bm{\Phi}}^{\text{MN}}=\frac{1}{2}\Phi_{\text{MN}}\int_{\bm{r}}\sum_{m_{1},m_{2}\in\mathbb{Z}}\hat{\bm{\Phi}}_{\varphi_{\text{MN}}}\cdot\bm{\phi}_{m_{1},m_{2}}(\bm{r})\,\\ \times  c_{\alpha}^{\dagger}(\bm{r}+\bm{R}_{m_{1},m_{2}})\,c_{\alpha}^{\:}(\bm{\bm{r}})+\text{H.c.},
\end{gathered}
\end{equation}
with multi-index $\alpha=\left(\sigma, \ell, s, \eta\right)$.
We further focus on the lowest moir\'e-lattice harmonic by setting $\phi_{m_{1},m_{2}}(\mathbf{r})=\phi_{m_{1},m_{2}}$ and only keeping the terms with the shortest possible $\mathbf{R}_{m_{1},m_{2}}$. Intuitively, MN order can be thought of as a distortion of the effective inter-moir\'e-unit-cell hopping matrix elements, as illustrated schematically in the lower right panel of \figref{fig:bandmoire}(d).

Conversely, GN acts as a local order parameter, $\Delta\vec{r}=0$ in \equref{eq:nematorder1}, without any explicit reference to the moir\'e scale, 
\begin{equation}
\label{eq:nematorder3}
    \mathcal{H}_{\bm{\Phi}}^{\mathrm{GN}}=\Phi_{\text{GN}}\int_{\bm{r}}\hat{\bm{\Phi}}_{\varphi_{\text{GN}}}\cdot\bm{\phi}_{\ell,s;\ell^{\prime},s^{\prime}}(\eta;\bm{r})\,c_{\ell,s}^{\dagger}(\bm{r})c^{\:}_{\ell^{\prime},s^{\prime}}(\bm{r}).
\end{equation}
Here, the correct transformation properties of $\vec{\phi}$ result from its structure in the internal indices. Focusing on the local intra-layer contributions and the leading (constant) basis function, the most general form reads as
\begin{equation}
\label{eq:nematorder4}
    \bm{\phi}_{\ell,s;\ell^{\prime}, s^{\prime}}(\eta;\vec{r})=\delta_{\ell,\ell^{\prime}}\psi_{\ell} \begin{pmatrix}
     (e^{i\alpha_{\ell}\eta \rho_{z}}\rho_{x})_{ss^{\prime}}\\
    \eta (e^{i\alpha_{\ell}\eta \rho_{z}}\rho_{y})_{ss^{\prime}}\\
  \end{pmatrix},
\end{equation}
where Pauli matrices in sublattice space are represented by $\rho_{j}$; $\alpha_{l}$ and $\psi_{l}$ are real-valued parameters. As shown schematically in the upper left panel of \figref{fig:bandmoire}(d), one can think of GN as the nematic distortion of the bonds of the individual graphene layers in a way that preserves the graphene translational symmetry.

We emphasize that GN and MN should not be viewed as distinct phases; they break the same symmetries and as such in general mix. We thus take $\mathcal{H}_{\bm{\Phi}}^{\text{MN}}+\mathcal{H}_{\bm{\Phi}}^{\mathrm{GN}}$ to describe nematicity in TDBG in the following, which depends on the set of parameters $\beta=\{\alpha_{\ell}, \psi_{\ell}, \Phi_{\text{MN}},\Phi_{\text{GN}}, \varphi_{\text{MN}}, \varphi_{\text{GN}}\}$. The computation of the LDOS for a specific set of parameters can be done straightforwardly from the continuum model. The resulting spatial dependence of the LDOS, $\mathcal{D}_{\omega_0}(\mathbf{r})$, is also shown in \figref{fig:bandmoire}(d) for two different values of $\beta$. As opposed to the plots without nematic order, $C_3$ is now broken, leading to stripes in the VFB, while translational symmetry is still preserved. The inverse problem---inferring the value of the parameters $\beta$ from a given LDOS pattern---is a much more challenging task. Our goal in the following sections will be to use ML, in particular, CNNs to learn the set $\beta$ directly from LDOS images.

\subsection{ML architecture}
\label{sec:mlarchitecture}
\begin{figure*} [htp!]
	\includegraphics[width=\textwidth]{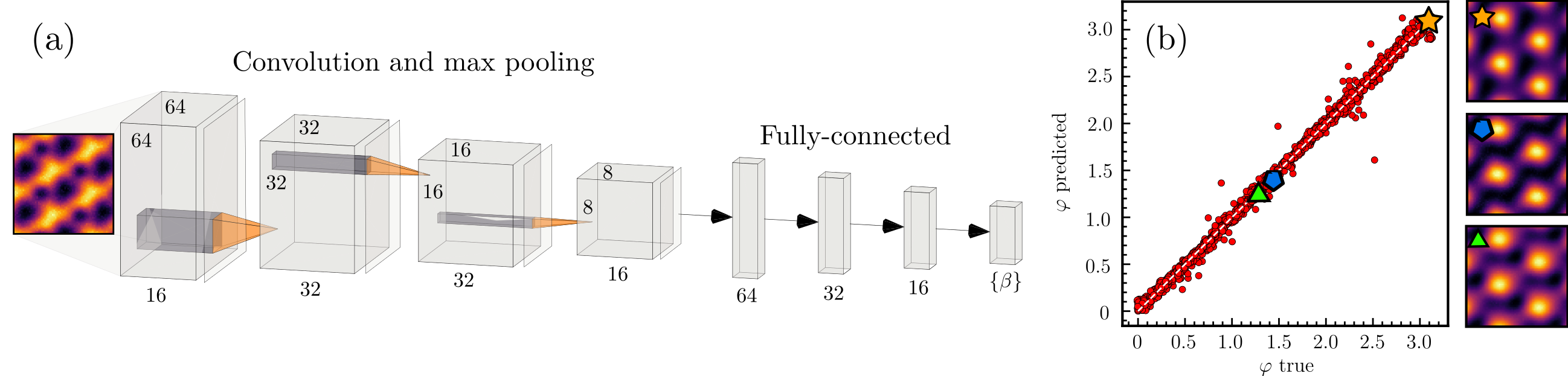}\caption{(a) Schematic figure of the CNN architecture used with only one $\mathcal{D}_{\omega_0}(\bm{r})$ input channel at an energy $\omega_0$ in the VFB, see \secref{sec:mlarchitecture} for details on the architecture and training dataset. (b) Comparison between true and predicted nematic director angles $\varphi$. Three samples of $\mathcal{D}_{\omega_0}(\bm{r})$ (star, pentagon and triangle) are displayed to emphasize that the relation between the LDOS and $\varphi$ is highly non-trivial as a result of the presence of different forms of nematicity.}
	\label{fig:nematicdirector}
\end{figure*}

Using CNNs to solve this inverse problem can be interpreted as a supervised learning task \cite{dawidModernApplicationsMachine2022}, i.e., a regression-like procedure using synthetic LDOS data labeled by their respective value of nematicity parameters $\beta$. 
More specifically, our CNNs take as inputs $65\times 65$ pixels of LDOS images and apply consecutive transformations (represented by a set of weights between each layer) in order to extract meaningful correlations that represent the set $\beta$. One example of the CNN image inputs is shown in \figref{fig:nematicdirector}(a). The complete dataset consists of 12000 images which are  divided into training ($60\%$), validation $(20\%)$ and test $(20\%)$ subgroups. Each image is generated for a randomly sampled set of nematic parameters $\beta$ and the intensities in the LDOS are modified with the addition of Gaussian noise (see \appref{app:DetailsModel}). The motivation for noise is twofold: to avoid overfitting \cite{goodfellowExplainingHarnessingAdversarial2015} and to test the stability against and performance of the procedure with noise, which is inevitably present in experimental data.

The ML architecture chosen for this task is represented in \figref{fig:nematicdirector}(a) and its implementation was done with the TensorFlow library \cite{tensorflow2015-whitepaper}. Each convolutional layer is followed by batch normalization and max pooling layers (Conv-Batch-MaxPool). The batch normalization layers normalize the input weights in each stage, and also reduce the number of epochs necessary for convergence \cite{ioffeBatchNormalizationAccelerating2015}. This process is repeated four times, with the convolutional layers having a kernel size of $3\times 3$ and strides set to $1$. The filters follow a sequence of $16-32-32-16$ with rectified linear unit (ReLU) activation functions \cite{fukushimaCognitronSelforganizingMultilayered1975}. Padding is set to zero such that the reduction of dimensionality is performed only by the MaxPool layers. In turn, these have both strides and pool sizes set to $2\times 2$. After a Flatten stage, dense layers lead to a dropout before the final layer with filters equal to the number of parameters in $\beta$. 
The Flatten layer transforms the data to a one-dimensional shape, and the Dropout reduces overfitting by setting a percentage of 20\% adjusted weights to zero \cite{srivastavaDropoutSimpleWay2014}. Tests on variations of this architecture are briefly described in \appref{app:ml}.

The learning procedure is then defined by the minimization of the loss function with respect to the CNN's weights in a backward propagation procedure \cite{rumelhartLearningRepresentationsBackpropagating1986}. The loss function can be represented as the mean squared error (MSE), which is defined as the difference between the true and expected set of parameters $\beta$ in $\text{MSE}=\sum_{j}^{N}\left(\beta_{j}^{\text{true}}-\beta_{j}^{\text{predicted}}\right)^{2}/N$, with $N$ representing the number of samples in the test dataset. Finally, we consider the adaptive moment estimation (ADAM) for the minimization of the loss function, with a learning rate of 0.001 and batch size equal to 64
\cite{kingmaAdamMethodStochastic2017}. After the completion of the training stage, the algorithm is ready to be deployed to previously unseen data, returning as outputs the parameters $\beta^{\text{predicted}}$.

\subsection{Orientation of nematic director}
\label{sec:orientation}
As a first investigation, we consider the task of predicting the orientation $\varphi$ of the nematic director from  $\mathcal{D}_{\omega_{0}}(\mathbf{r})$ images at a single energy in the VFB [$\omega_{0}=-15$ meV, see \figref{fig:bandmoire}(b)]. For this, we consider a dataset with randomly generated MN and GN intensities $\Phi^{\text{MN}}, \Phi^{\text{GN}}\in[0.001,0.1]$~eV, and $\varphi^{\text{MN}}=\varphi^{\text{GN}}=\varphi\in [0,\pi]$. Furthermore, $\psi_{l}=1$ and $\alpha_{l}=0$ for all layers. The relation between the shape of the LDOS at a single energy $\mathcal{D}_{\omega_{0}}(\mathbf{r})$ and $\varphi$ is highly non-trivial for two reasons: even for a given form of nematicity, changing $\varphi$ generically not just merely rotates the LDOS pattern, due to the lattice, but leads to complex distortions of its structure. Additionally, by sampling $\mathcal{H}_{\bm{\Phi}}^{\text{MN}}+\mathcal{H}_{\bm{\Phi}}^{\mathrm{GN}}$, even if the same bond direction is favored over the $C_{3}$-related ones in the LDOS pattern of two samples, the underlying $\varphi$ can be rather different. As can be seen in the three sample LDOS plots in \figref{fig:nematicdirector}(b) with different values of $\varphi$, the correspondence between $\varphi$ and $\mathcal{D}_{\omega_{0}}(\mathbf{r})$ is complex and not apparent to the human eye.

Using the angles $\varphi$ as labels to the data is the most straightforward choice, but leads to inaccurate predictions around $0$ and $\pi$ due to the periodicity in the definition of the nematic order parameter, $\hat{\vec{\Phi}}_{\varphi}=(\cos2\varphi, \sin2\varphi)=\hat{\vec{\Phi}}_{\varphi+\pi}$. 
To circumvent this feature, we use the two-component label $\hat{\vec{\Phi}}_{\varphi}$ instead of $\varphi$ in the training process and then fold the network's prediction back to $\varphi$ with the $arctan2$ function \cite{fischerImageOrientationEstimation2015}. The results, shown in \figref{fig:nematicdirector}(b), are consistent with the true labels, including at the boundaries of $\varphi$'s domain. This shows that even when the precise nature of nematicity (predominantly MN or GN or an admixture of the two) is not known, the director orientation $\varphi$ can be accurately predicted with our CNN setup from $\mathcal{D}_{\omega_{0}}(\mathbf{r})$ at a single energy. We have checked that the few outliers in \figref{fig:nematicdirector}(b) are directly related to small nematic intensities, where $\varphi$ has virtually no impact on the LDOS and is, thus, impossible to predict. 

\subsection{Form of nematicity}
\label{sec:formofnematicity}
After successfully learning the director orientation $\varphi$ in the presence of different nematicities, we proceed into investigating finer details of these couplings by learning the parameters $\beta=\{\Phi^{\text{MN}}, \Phi^{\text{GN}}, \alpha_{l}\}$ defined in Eqs.~(\ref{eq:nematorder2}-\ref{eq:nematorder4}). To this end, we consider $\psi_{l}=1$ and $\alpha_{l}=\alpha$ for all layers. For concreteness, we set $\varphi^{\text{MN}}=\varphi^{\text{GN}}=\varphi=2\pi/3$, which is one of the possible discrete orientations ($\varphi^{\text{MN}}=\varphi^{\text{GN}}=2\pi/3, \pi/6$ and symmetry related) of the nematic director in presence of $C_{2x}$. 
The dataset now consists of randomly generated MN and GN intensities $\Phi^{\text{MN}}, \Phi^{\text{GN}}\in[0.001,0.1]$ eV, and $\alpha\in [0,\pi]$. The intensity values are chosen such that the stripes in the VFB resemble the experimental results \cite{rubio-verduMoireNematicPhase2022}. As with $\varphi$, instead of learning the angular variable $\alpha$ directly, the $arctan2$ mapping from Section \ref{sec:orientation} is applied.

\begin{figure}
	\centering{}\includegraphics[width=\columnwidth]{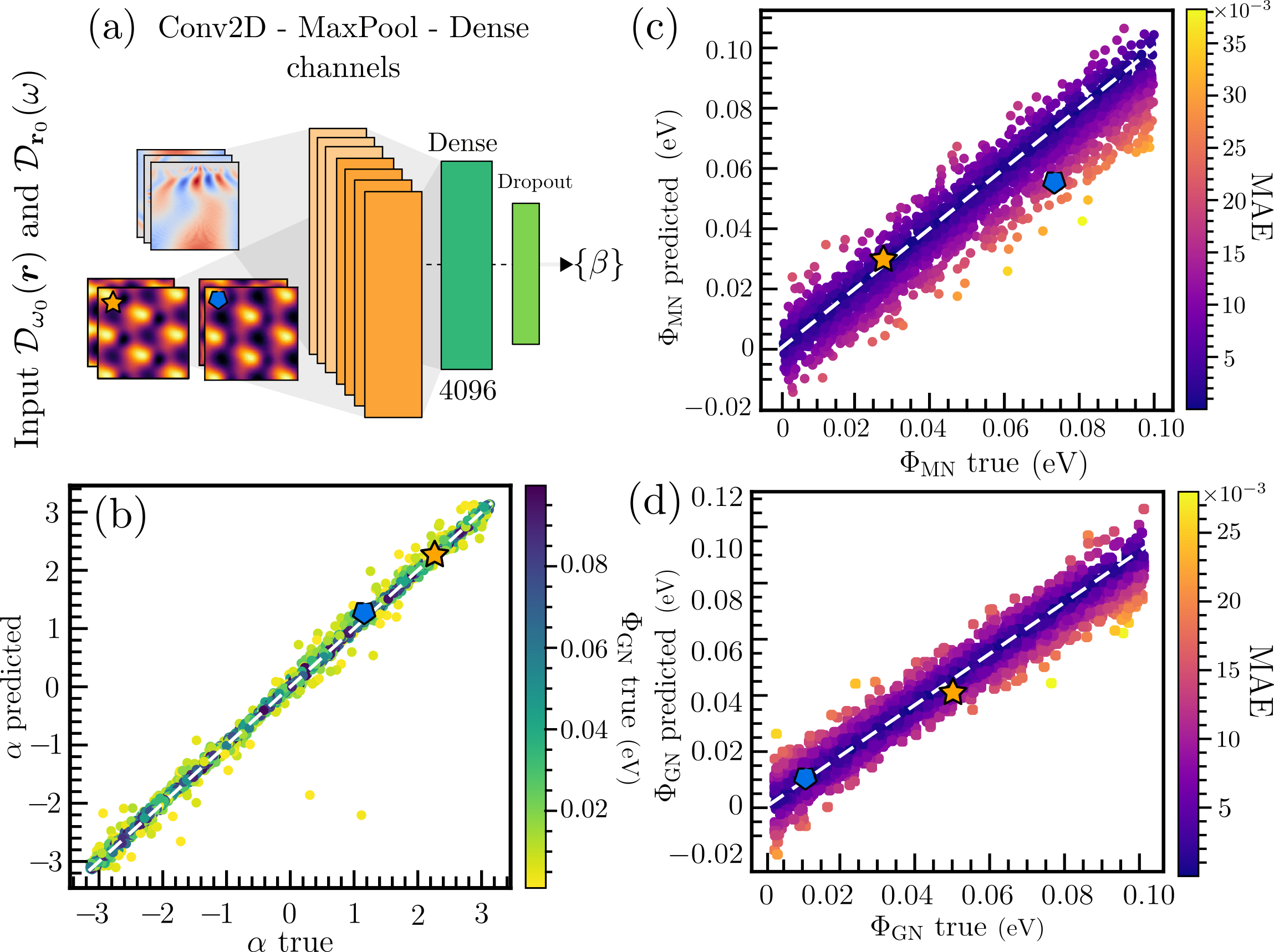}\caption{(a) CNN architecture used for learning the nematic microscopic parameters. Each 'Conv2D-MaxPool-Dense' channel refers to the structure from \figref{fig:nematicdirector}(a). (b) Predicted versus true $\alpha$ parameter, with outliers (brighter colors) being related to small graphene nematic intensity $\Phi_{\text{GN}}$. (c-d) Predicted versus true parameters for graphene and moiré intensities, with colorbars representing the mean absolute error (MAE) in the intensities. Two examples (star and pentagon) indicate that two very different forms of nematicity can lead to very similar LDOS patterns at a single energy, making the inclusion of several channels necessary.}
	\label{fig:nematicform}
\end{figure}

Using only the LDOS at a single energy (i.e.~one $\mathcal{D}_{\omega_0}(\mathbf{r})$ channel) in the ML architecture for this task does not produce accurate predictions. Additionally, both hyperparameter optimization and architecture modifications did not lead to any significant improvement, implying that nematic order impacts the electronic structure in complex ways that cascade across energy scales. In fact, this is also intuitively clear since, for example, the samples marked by a star and pentagon in \figref{fig:nematicform}(a) have fundamentally different nematic couplings and yet  exhibit visually similar $\mathcal{D}_{\omega_0}(\mathbf{r})$ images at the VFB energy.

In experiments, one can typically obtain single point spectra [$\mathcal{D}_{\vec{r}_0}(\omega)$] and real space LDOS images at fixed energies [$\mathcal{D}_{\omega_0}(\mathbf{r})$]. We can therefore include additional input channels corresponding to $\mathcal{D}_{\omega_0}(\mathbf{r})$ and $\mathcal{D}_{\vec{r}_0}(\omega)$ for different energies $\omega_{0}$ and points $\vec{r}_0$, respectively. In the second case, the individual point spectra are transformed to scaleogram images for consistency with the input data for CNNs \cite{berthusenLearningCrystalField2021,mallatWaveletTourSignal1999}, see upper left inset in \figref{fig:nematicform}(a) and \appref{app:DetailsModel}. The new architecture is then formed by four channels with $\mathcal{D}_{\omega_0}(\mathbf{r})$  inputs at fixed energies $\omega_{0}=\left(-35, -15, 1, 23\right)$ meV within the flat and remote bands, such that they resemble visually the corresponding ones in the experimental data of \refcite{rubio-verduMoireNematicPhase2022},
and three channels for $\mathcal{D}_{\vec{r}_0}(\omega)$  scaleogram inputs at stacking positions $\vec{r}_0=\{\text{ABAB, BAAC, ABCA}\}$, cf.~\figref{fig:bandmoire}(c). Each channel is passed through parallel Conv-Batch-MaxPool layers as in \figref{fig:nematicdirector}(a), but instead of flattening each channel separately, they are concatenated to a Dense-Dropout stage before the last layer [\figref{fig:nematicform}(a)]. 

In \figref{fig:nematicform}(b-d), predictions on the test data set are represented for (b) $\alpha$, and (c) the moir\'e and (d) graphene nematic intensities; as can be seen, very good agreement is found between the reconstructed and true parameters.
The outliers in $\alpha $  are related to small $\Phi_{\text{GN}}$ (brighter colors). From \equsref{eq:nematorder3}{eq:nematorder4}, it is clear that for small $\Phi_{\text{GN}}$, minimal changes will be induced in the LDOS, irrespective of the true value of the phase governed by $\alpha$. This is a similar behavior to what was observed for outliers in the nematic director prediction. 
If $\alpha$ is maintained fixed, we observed (not shown) that predictions for $\Phi_{\text{GN}}$ and $\Phi_{\text{MN}}$ get more accurate. 
The results of \figref{fig:nematicform} demonstrate that the microscopic form of nematicity can be extracted from the LDOS if significant energy dependence is included  in the input data set.
 
\subsection{Including strain}
\label{sec:includingstrain}
As already alluded to above, another possible source of $C_3$ breaking is strain \cite{nguyenStraininducedModulationDirac2015,yanStrainCurvatureInduced2013, 
huderElectronicSpectrumTwisted2018,biDesigningFlatBands2019}, which is believed to be a ubiquitous property of graphene moir\'e superlattices at small twist angles. Breaking the same symmetries as nematic order, strain can obscure the experimental identification of nematic order and their precise interplay is still under debate \cite{kerelskyMaximizedElectronInteractions2019,jiangChargeOrderBroken2019,choiElectronicCorrelationsTwisted2019,scheurerSpectroscopyGrapheneMagic2019}. Experiments indicate \cite{huderElectronicSpectrumTwisted2018,kerelskyMaximizedElectronInteractions2019,jiangChargeOrderBroken2019,choiElectronicCorrelationsTwisted2019,rubio-verduMoireNematicPhase2022} that the most relevant form of strain in graphene superlattices such as twisted bilayer graphene or TDBG is uniaxial heterostrain. In this case, the matrices $\mathcal{E}_j$ describing the in-plane metric deformation of the coordinates in the $j$th rotated bernal bilayer of TDBG are of the form 
\begin{equation}
\mathcal{E}_2 = -\mathcal{E}_1 = \frac{1}{2}R(\theta_{\epsilon})^{-1}\begin{pmatrix}
    -\epsilon & 0\\
    0 & \nu \epsilon \\
  \end{pmatrix} R(\theta_{\epsilon}). \label{DefinitionOfStrain}
\end{equation}
Here $\nu=0.16$ is the Poisson ratio for graphene and $R(\theta_{\epsilon})$ is the $2\times 2$ matrix describing rotations of 2D vectors by angle $\theta_{\epsilon}$. We see that uniaxial heterostrain is characterized by two variables, the strain intensity $\epsilon$ and the direction of strain, parameterized by the angle $\theta_{\epsilon}$. 

In the following, we allow for the simultaneous presence of uniaxial heterostrain and nematic order, leading to two additional parameters, $\epsilon$ and $\theta_{\epsilon}$, in $\beta$. We will study whether our ML approach is still able to extract the microscopic form of nematicity and also learn the relative strength and direction of strain. Note that the form of nematicity is still given by Eqs.~(\ref{eq:nematorder2}-\ref{eq:nematorder4}), with the only difference that we replace $\mathbf{L}_{j}^{M}$ in the definition of $\mathbf{R}_{m_{1},m_{2}}$ by the strained moir\'e lattice vectors.
The data set for this task is built with nematic intensities $\Phi^{\text{MN}}, \Phi^{\text{GN}}\in[0.001,0.1]$ eV, with the addition of strain parameters $\epsilon \in [0,0.8]\:\% $ and $\theta_{\epsilon}\in[0,\pi/3]$. Here, $\alpha_{l}=0$, $\psi_{l} = 1$ and $\varphi=\varphi_{\text{MN}}=\varphi_{\text{GN}}=2\pi/3$. The domain for the strain intensities is chosen based on typical values observed in TBG \cite{kerelskyMaximizedElectronInteractions2019}, and for $\theta_{\epsilon}$ on the periodicity of the unstrained system as $\theta_{\epsilon}\rightarrow \theta_{\epsilon}+\pi/3$ \cite{biDesigningFlatBands2019}. The ML architecture employed in this section is the same as in the previous investigation [\figref{fig:nematicform}(a)]. 

\begin{figure}
\includegraphics[width=\columnwidth]{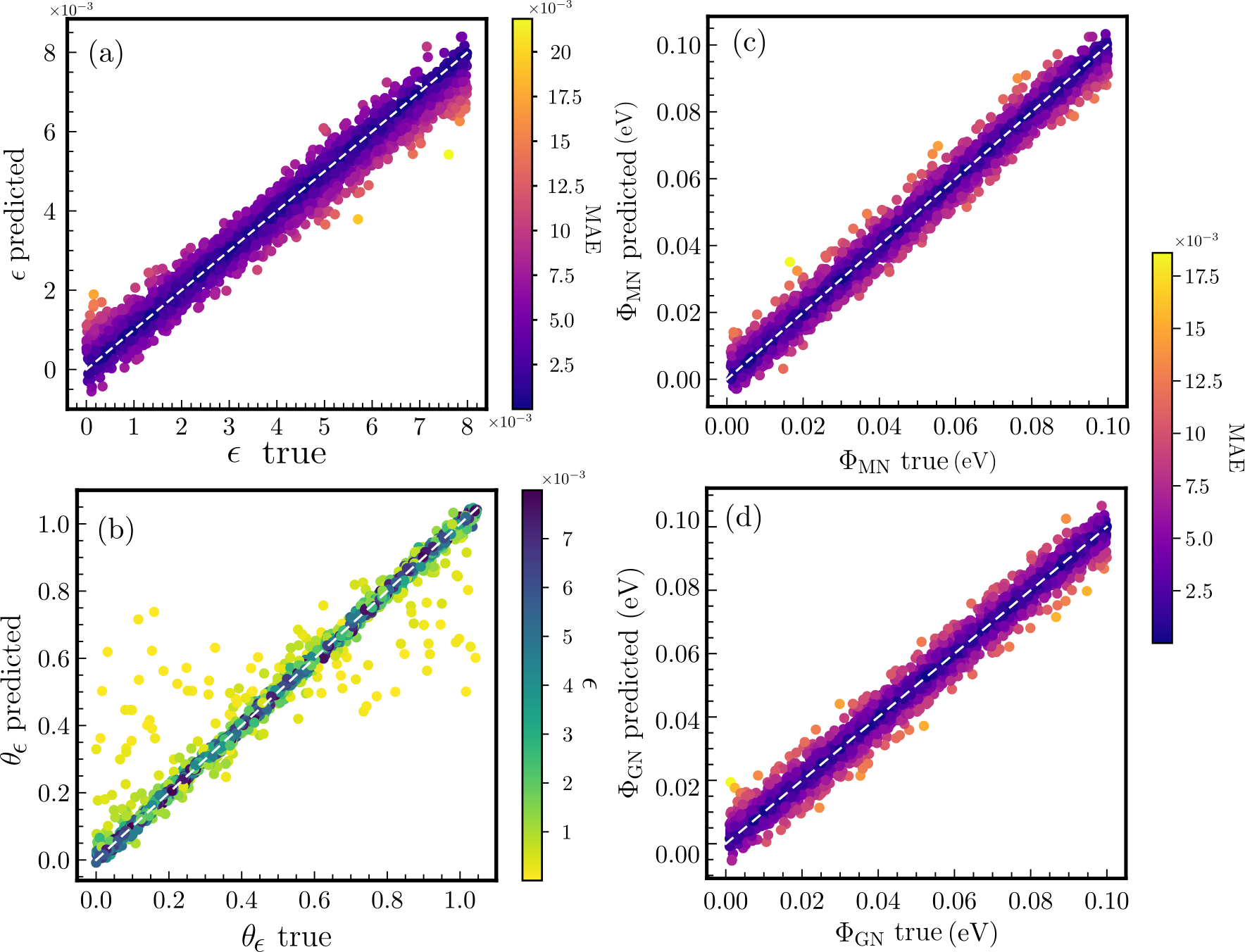}\caption{Predicted versus true values for the strain intensity $\epsilon$ (a) and  angle $\theta_{\epsilon}$ (b). The prediction for the nematic intensities are depicted in panels (c) and (d). The CNN architecture used to produce these results is described in Fig.~\ref{fig:nematicform}(a). Similarly to the prediction of the $\alpha$ parameter in the presence of only nematicity, outliers in $\theta_{\epsilon}$ are related to small $\epsilon$.} 
 \label{fig:strainandnematic}
\end{figure}

In \figref{fig:strainandnematic}(a-d), predictions on the test data set are shown for $\epsilon$ (a), $\theta_{\epsilon}$ (b), and the nematic intensities (c-d). 
At first sight, the result for the strain angle in \figref{fig:strainandnematic}(b) looks as if the procedure ceased to work, since there are many data points where the true and predicted value of $\theta_{\epsilon}$ differ significantly. However, when indicating the true strain intensity label $\epsilon$ for each prediction, it becomes clear that the outliers are related to small values of $\epsilon$ (brighter colors). As such, this behavior is not a shortcoming of the learning procedure but actually a feature of strain: for small enough $\epsilon$ in \equref{DefinitionOfStrain}, the angle $\theta_{\epsilon}$ has no meaning. We have checked that removing the samples with small strain $\epsilon$ from the training and test data set will lead to accurate predictions of $\theta_{\epsilon}$. The stability that we find for our learning procedure in the presence of virtually vanishing $\epsilon$ is, however, important when applying it to experimental data, where the strength of strain is unknown.

Most importantly, we see in \figref{fig:strainandnematic}(c-d) that the nematic couplings can still be accurately predicted when varying strain is present. The MAE is equally distributed in these cases, in contrast to the strain intensity prediction. 
This shows that not only nematic order can be identified when strain is present, but also its internal structure and the strength of strain that is present at the same time can be resolved when using different channels consisting of both $\mathcal{D}_{\vec{r}_0}(\omega)$ and $\mathcal{D}_{\omega_0}(\mathbf{r})$ as inputs. This allows the networks to take into account correlations between different energies in the STM data, which in turn conveys the crucial microscopic physics, enabling the model to disambiguate between lattice and electronic effects. 

\subsection{Experimental data}
\label{sec:experimentaldata}

\begin{figure} [ht]
\includegraphics[width=\columnwidth]{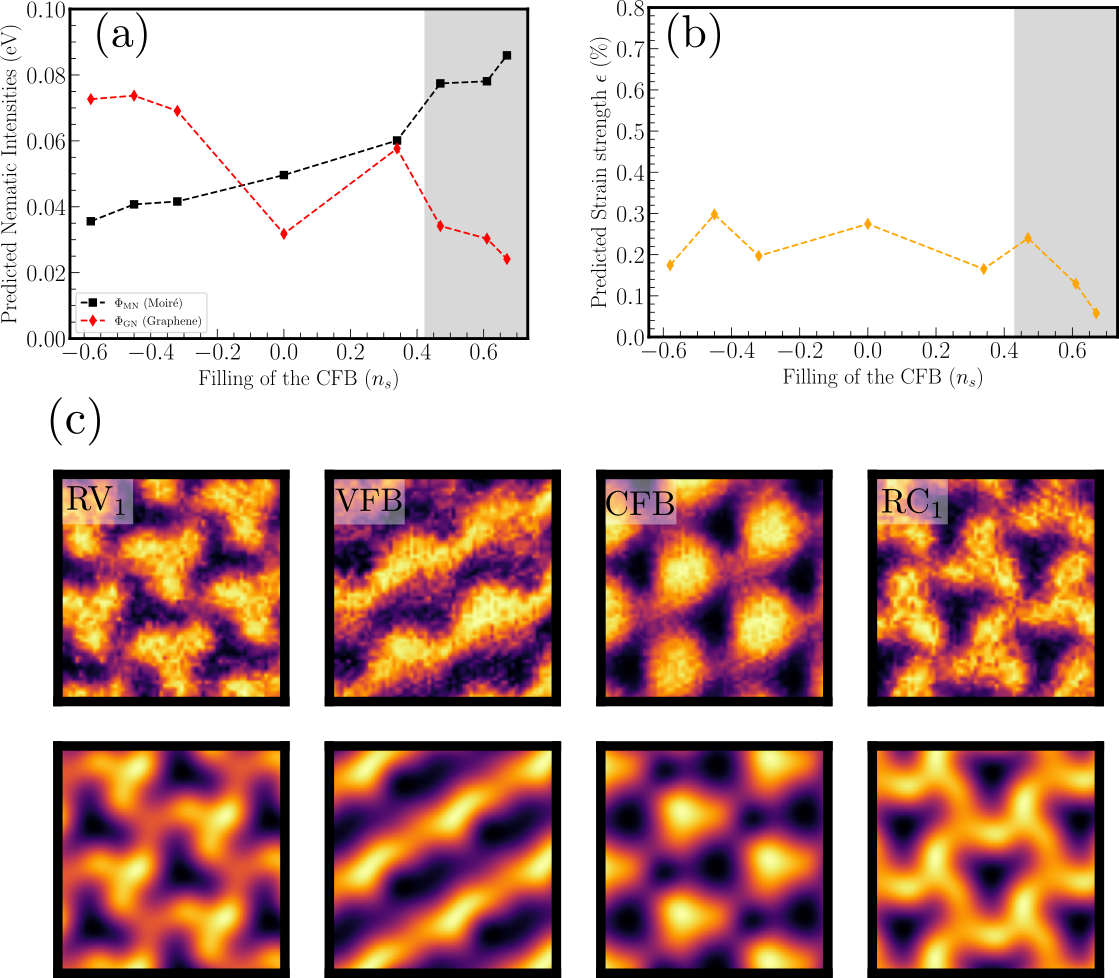}\caption{Predicted values from the trained CNN to nematic intensities (a) and strain strength (b) as a function of the filling of the CFB ($n_{s}$). The gray region ($n_{s}\ge0.47$) indicates the fillings where the continuum model showed more resemblance to the experimental data obtained in \refcite{rubio-verduMoireNematicPhase2022}. In panel (c) the experimental $\mathcal{D}_{\mathbf{r}_{0}} \left(\omega\right)$ channels for $n_{s}=0.67$ are shown for comparison with the ones obtained from the continuum model with the parameters $\beta_{\text{exp}}=\{\Phi_{\text{MN}}, \Phi_{\text{GN}}, \epsilon\}=\{0.086\,\text{eV}, 0.024\, \text{eV}, 0.05\%\}$ predicted by the trained CNN.} 
\label{fig:expprediction} 
\end{figure}

After demonstrating the effectiveness of CNNs on learning microscopic parameters $\{\beta_{i}\}$ from a synthetic (theoretical) data set $D_{\text{th}}\left(\beta_{1},\cdots, \beta_{N_{\text{th}}}\right)$ with $N_{\text{th}}$ samples, 
we now proceed into applying the trained ML architecture for predictions of the \textit{a priori} unknown sets of parameters $\{\beta^{\prime}_{i}\}$ in an experimental data set $D_{\text{exp}}\left(\beta_{1}^{\prime},\cdots, \beta_{N_{\text{exp}}}^{\prime}\right)$. 
For concreteness, we use the same synthetic training data set as in \appref{app:ml}, where only the nematic and strain intensities are predicted, i.e., $\beta=\{\Phi_{\text{MN}}, \Phi_{\text{GN}}, \epsilon\}$.
The data set $D_{\text{exp}}$ is constituted of both scaleograms $\mathcal{D}_{\mathbf{r}_{0}}(\omega)$ and $\mathcal{D}_{\omega_{0}}(\mathbf{r})$ maps for different fillings of the CFB ($n_{s}$). 
More details about the preprocessing of the experimental data $D_{\text{exp}}$ can be found in \appref{app:expdata}.

In \figref{fig:expprediction}, predictions of the trained CNN for the set $\{\beta_{i}^{\prime}\}$ show non-zero values of nematicity (a) and strain (b) for all fillings of the CFB. For $n_{s}\ge0.47$ (gray region), the experimental data shows the most pronounced signatures of broken rotational symmetry to the human eye, which was previously interpreted as electronic nematic order \cite{samajdarElectricfieldtunableElectronicNematic2021,rubio-verduMoireNematicPhase2022}. Here the CNN predicts MN to dominate over GN, although both are finite (as expected by symmetry). As can be seen in \figref{fig:expprediction}(c), the parameters predicted by the CNN nicely reproduce the key features in the experimental data, including the strong stripes in the VFB and the much weaker, albeit finite, signatures of nematicity in the other bands. 

For smaller fillings, $n_{s}<0.47$, the experimental data still exhibit distortions that break $C_3$, see \appref{app:expdata}, but no clear stripe-like features appear. The CNN tries to assign different anisotropy sources to these distorted regions, but the agreement between theoretical prediction and experiment is less accurate than for larger $n_s$. It is clearly possible that, indeed, a crossover from primarily MN to GN occurs when lowering $n_s$, as predicted by the neural network, see \figref{fig:expprediction}(a), in particular, since nematic order is also a plausible instability in non-twisted bilayer graphene \cite{liuTuningElectronCorrelation2021, cvetkovicElectronicMulticriticalityBilayer2012}.
However, we believe that additional experimental data and refined theoretical models are required to conclude whether this is really the case.

In contrast to this interplay between the nematic couplings, strain remains relatively constant for all $n_s$, and slightly decreases in \figref{fig:expprediction}(b) for $n_{s}\ge 0.47$ as it approaches the same order of magnitude of $\epsilon \in [0.003-0.1\%]$ that is expected for the experimental samples in $D_{\text{exp}}$ \cite{rubio-verduMoireNematicPhase2022}. We note that at low fillings the value of strain that is predicted by the neural network is nevertheless significantly greater than the value extracted from experimental topography.  This is likely a consequence of subtle differences between the continuum model calculations and the experimental spectroscopy, which the network attempts to accommodate by including finite strain.   

\section{Discussion}
We constructed and demonstrated a ML procedure that can extract the form of the nematic order parameter in TDBG from LDOS data. The key ingredient was the use of several channels that capture the correlations among different energies. Our work has several important implications. First, it shows that the presence and even the strength and internal structure of nematic order can be extracted when the sample exhibits significant heterostrain; this is a crucial aspect for moiré systems where the issue of distinguishing between nematicity and strain has been subject of debate. Second, our analysis also shows which type of STM data is needed and most useful to extract information about nematicity: as we have seen, the LDOS maps at a single energy, $\mathcal{D}_{\omega_0}(\vec{r})$, are not enough to deduce the form of the nematic order parameter and---contrary to what one might have expected---point spectra, i.e., $\mathcal{D}_{\vec{r}_0}(\omega)$, contain a lot of helpful complementary information for that task (see also the second model discussed in \appref{app:tbgnematic}). We emphasize that this form of `solid-state Hamiltonian learning’, i.e., of parameterizing the leading terms of a set of microscopic order parameters (like nematic order) or perturbations (such as strain) and extracting their form using multi-channel CNNs can be more broadly applied---to other systems, see \appref{app:tbgnematic} where we discuss a toy model for twisted-bilayer graphene, and other forms of instabilities.
As such, this could open up completely new ways of revealing the form and role of nematic order and other phases for the physics of quantum materials.

\begin{acknowledgments}
J.A.S.~and M.S.S.~acknowledge funding by the European Union (ERC-2021-STG, Project 101040651---SuperCorr). Views and opinions expressed are however those of the authors only and do not necessarily reflect those of the European Union or the European Research Council Executive Agency. Neither the European Union nor the granting authority can be held responsible for them. Salary support is also provided by the National Science Foundation via grant DMR-2004691 (S.T.) and by the Office of Basic Energy Sciences, Materials Sciences and Engineering Division, U.S. Department of Energy under Contract No. DE-SC0012704 (A.N.P.). J.A.S. is grateful for discussions with J. P. Valeriano, Sayan Banerjee, Patrick Wilhelm, Igor Reis and Pedro H. P. Cintra. M.S.S.~also thanks R.~Samajdar, R.~Fernandes, and J.~Venderbos on a previous collaboration on nematic order in TDBG \cite{samajdarElectricfieldtunableElectronicNematic2021}.
\end{acknowledgments}

\bibliography{ML_Nematicity_TDBG}

\onecolumngrid

\begin{appendix}

\section{Continuum model and LDOS maps}\label{app:DetailsModel}

Twisted double-bilayer graphene (TDBG) consists of two bilayer graphene stacks with primitive lattice vectors $\bm{a}_{1}=a\left(1, 0\right)$ and $\bm{a}_{2}=a\left(1, \sqrt{3}\right)/2$, where $a\approx 0.246$ nm is the lattice constant of graphene, and corresponding reciprocal lattice vectors $\vec{b}_j$, following from 
$\bm{a}_{i}\cdot\bm{b}_{j}=2\pi\delta_{ij}$. After applying a twist angle $\theta$ between the Bernal stacks, these vectors are modified by the two-dimensional rotation matrix $R(\theta)$ as $\left(\bm{a}^{l}_{i},\bm{b}^{l}_{i}\right)=R\left(\mp\theta/2\right)\left(\bm{a}_{i},\bm{b}_{i}\right)$ with $-,+$ for each stack $l=1,2$. The corresponding Dirac cones in each valley $\eta=\pm 1$ of the individual graphene layers are located at $\bm{K}^{l}_{\eta}=-\eta \left(2\bm{b}_{1}^{l}+\bm{b}_{2}^{l}\right)/3$. The emerging moiré pattern [with triangular Bravais lattice shown as green domains in \figref{fig:bandmoire}(a)] is represented by the difference of the new lattice vectors from each bilayer stack in reciprocal space as $\bm{G}_{i}^{M}=\bm{b}_{i}^{1}-\bm{b}_{i}^{2}\:\left(i=1,2\right)$, with corresponding primitive lattice vectors $\bm{L}_{j}^{M}$ obtained from the relation $\bm{G}_{i}^{M}\cdot\bm{L}_{j}^{M}=2\pi \delta_{ij}$.

We consider a description of the low-energy physics for TDBG  
via the continuum Hamiltonian of \refcite{koshinoBandStructureTopological2019}. In the Bloch basis given by carbon's p$_{z}$ 
orbitals $\left(A_{1},B_{1},\dots,A_{4},B_{4}\right)$, with sublattices $s=\{A_{\ell}, B_{\ell}\}$
and layers $\ell=1,2,3,4$, the continuum Hamiltonian in valley $\eta$ for small
twist angles ($\theta \ll 1$) in AB-AB double bilayer graphene can be written as  
\begin{equation}
	H_{AB-AB} = \begin{pmatrix}
	 H_{0}\left(\boldsymbol{k}_{1}\right) & S^{\dagger}\left(\boldsymbol{k}_{1}\right) & & \\
	 S\left(\boldsymbol{k}_{1}\right) & H_{0}^{\prime}\left(\boldsymbol{k}_{1}\right) & U^{\dagger} & \\
	 & U & H_{0}\left(\boldsymbol{k}_{2}\right) &S^{\dagger}\left(\boldsymbol{k}_{2}\right)  \\
	 & & S\left(\boldsymbol{k}_{2}\right) &  H_{0}\left(\boldsymbol{k}_{2}^{\prime}\right) \\
	\end{pmatrix} + V
	\label{eq:cm}
\end{equation}
with Bloch wavevectors generated by
$\boldsymbol{k}_{j}= R(\mp \theta/2)\left(\boldsymbol{k}-
  \boldsymbol{K}_{\eta}^{j}\right)\left(j=1,2\right)$ and single-layer graphene Hamiltonians with $k_{\pm}=\eta k_{x}\pm ik_{y}$ as
\begin{equation}
 H_{0}\left(\boldsymbol{k}\right) = \begin{pmatrix}
  0 & -\hbar \nu k_{-}\\
   -\hbar \nu k_{+} & d \\
 \end{pmatrix}\quad \text{and} \quad H_{0}^{\prime}\left(\boldsymbol{k}\right) = \begin{pmatrix}
  d & -\hbar \nu k_{-}\\
   -\hbar \nu k_{+} & 0 \\
 \end{pmatrix}.
  \label{eq:h0}
\end{equation}
The explicit matrix structure in \equref{eq:h0} refers to sublattice space (associated with Pauli matrices $\rho_j$ in the main text).
In turn, $H_0$ and $H_0'$ are coupled via 
\begin{equation}
S\left(\boldsymbol{k}\right) = \begin{pmatrix}
  \hbar \nu_{4}k_{+} & \gamma_{1}\\
   \hbar \nu_{3}k_{-} & \hbar \nu_{4}k_{+} \\
 \end{pmatrix},
  \label{eq:sk}
\end{equation}
with parameters $\{d, \hbar \nu/a,\gamma_{1}, \nu_{3}, \nu_{4}\}=\{0.050, 2.776, 0.4, 
0.32, 0.044\}\: \text{eV}$ and $\nu_{i}=\left(\sqrt{3}/2\right)\gamma_{i}a/\hbar \left(i=3,4\right)$. For more details about the physical significance of each term, see Ref. \cite{koshinoBandStructureTopological2019}.
Considering a self-consistently calculated screened electric field, the  interlayer potential matrix reads as
\begin{equation}
V = \text{diag}\left(\Delta_{1}\rho_0,\Delta_{2}\rho_0,\Delta_{3}\rho_0,\Delta_{4}\rho_0\right)
\end{equation}
where $\rho_0$ is the unit matrix in sublattice space.
For a filling fraction of $\nu=0.475$, on-site potentials representing the electrostatic energy between adjacent layers are given by $\Delta=(\Delta_{1}, \Delta_{2}, \Delta_{3}, \Delta_{4})=(4.079,1.021,-1.537,-3.563)$ meV \cite{samajdarElectricfieldtunableElectronicNematic2021}, which we use in our numerical calculations.  

Finally, the moiré interlayer coupling defined between the twisted layers $\ell=2-3$ is given by
\begin{equation}
 U = \begin{pmatrix}
  u & u^{\prime} \\
  u^{\prime} & u \\
 \end{pmatrix} + \begin{pmatrix}
  u & u^{\prime}\omega^{-\eta} \\
  u^{\prime}\omega^{\eta} & u \\
 \end{pmatrix}e^{i\eta\boldsymbol{G}^{M}_{1}\cdot\boldsymbol{r}}  +
\begin{pmatrix}
  u & u^{\prime}\omega^{\eta} \\
  u^{\prime}\omega^{-\eta} & u \\
 \end{pmatrix}e^{i\eta\left(\boldsymbol{G}^{M}_{1}+
 \boldsymbol{G}^{M}_{2}\right)\cdot\boldsymbol{r}} \label{eq:moiréinterlayerhopping}
\end{equation}
with $\omega=\exp2\pi i/3$, $u=0.0797$ eV and $u^{\prime}=0.0975$
 eV. These parameters were chosen in accordance with Ref. \cite{samajdarElectricfieldtunableElectronicNematic2021, koshinoBandStructureTopological2019}.

From the continuum Hamiltonian in \equref{eq:cm}, a numerical diagonalization in momentum space is performed by selecting a finite number of $\bm{q}$ wavevectors in a cutoff circle $|\bm{q}-\bm{q}_{0}|<q_{c} $, with radius $q_{c}=4|\bm{G}_{i}^{M}|$ around the midpoint $\bm{q_{0}}=\left(\bm{K}_{\eta}^{1}+\bm{K}_{\eta}^{2}\right)/2$ between Dirac cones $\bm{K}_{\eta}^{j}$. Here, the Bloch vector $\mathbf{k}$ in the moiré Brillouin zone is hybridized with the graphene eigenstates at $\bm{q}=\bm{k}+\mathbf{G}_{n_{1},n_{2}}$ due to the coupling between Bernal bilayers via \equref{eq:moiréinterlayerhopping}, with $\mathbf{G}_{n_{1},n_{2}}=n_{1}\mathbf{G}_{1}^{M}+n_{2}\mathbf{G}_{2}^{M} \left(n_{1}, n_{2} \in \mathbb{Z}\right)$. Since we do not consider the intervalley graphene nematicity \cite{samajdarElectricfieldtunableElectronicNematic2021}, the calculations are performed with a fixed valley index, e.g., $\eta=+1$. The corresponding band structure for $\eta=-1$ can be obtained by a time-reversal symmetry transformation. 
 For a certain band $n$, the wave functions, truncated up to a wavevector $\mathbf{G}_{n_{1},n_{2}}^{c}$  in the reciprocal lattice, are represented as
 \noindent
 \begin{equation}
     \Psi_{n}\left(\mathbf{k}\right)=\left(\psi_{n,\mathbf{k}}\left(\mathbf{G}^{1}_{n_{1},n_{2}}\right), \cdots, \psi_{n,\mathbf{k}}\left(\mathbf{G}^{c}_{n_{1},n_{2}}\right)\right)^{T}
\end{equation}
with each term
\begin{equation}
 \psi_{n,\mathbf{k}}\left(\mathbf{G}\right)=\left(U^{A_{1}}_{n,\mathbf{k}}\left(\mathbf{G}\right),U^{A_{2}}_{n,\mathbf{k}}\left(\mathbf{G}\right),\cdots, U^{B_{4}}_{n,\mathbf{k}}\left(\mathbf{G}\right)\right)^{T}
 \end{equation}
 \noindent
 containing elements in layer $\ell$ and sublattice $s$ spaces. From these wave functions, the LDOS mappings $\mathcal{D}_{\vec{r}_0}(\omega)$ and $\mathcal{D}_{\omega_0}(\mathbf{r})$ are computed from 
\begin{equation}\mathcal{D}\left(\boldsymbol{r}, \omega\right)=\sum_{n,\boldsymbol{k}}
  \sum_{\boldsymbol{G}, \boldsymbol{G}^{\prime}}e^{-i\left(\boldsymbol{G}-\boldsymbol{G}^{\prime}\right)\vec{r}}\delta \left(\omega-\omega_{n,\boldsymbol{k}}\right)
  \times \left(\left[U^{A_{4}}_{n,\boldsymbol{k}}\left(\boldsymbol{G}^{\prime}\right)\right]^{*}U^{A_{4}}_{n,\boldsymbol{k}}\left(\boldsymbol{G}\right)
  +\left[U^{B_{4}}_{n,\boldsymbol{k}}\left(\boldsymbol{G}^{\prime}\right)\right]^{*}U^{B_{4}}_{n,\boldsymbol{k}}\left(\boldsymbol{G}\right)\right).
\end{equation}
where $\omega_{n,\boldsymbol{k}}$ is the corresponding eigenvalue to the wave function $\Psi_{n}\left(\boldsymbol{k}\right)$.
This is already projected onto the topmost graphene layer $\ell=4$, where tunneling of electrons from the STM tip are expected to occur in the experimental setup. 

\begin{figure}[bt]
	\centering{}\includegraphics[width=\columnwidth]{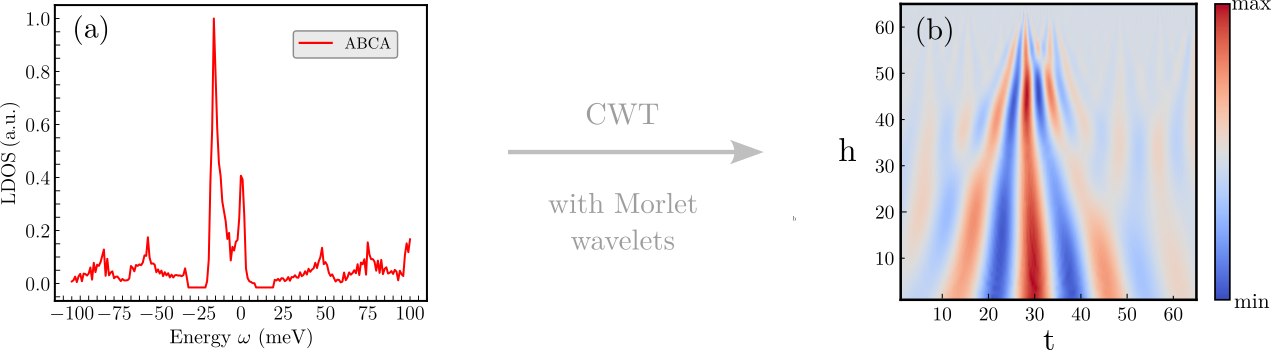}\caption{Local density of states for a fixed position $\mathcal{D}_{\vec{r}_0=\text{ABCA}}(\omega)$ as a function of energy $\omega$ (a) and its corresponding image  $W\left(t, h\right)$ after a continuous wavelet transformation (b). Intensities displayed are in arbitrary units.}
	\label{fig:wavelettransform}
\end{figure}
To transform the one-dimensional map of LDOS as a funtion of energies $\mathcal{D}_{\vec{r}_0}(\omega)=\{\mathcal{D}(\omega_{0}), \cdots, \mathcal{D}(\omega_{N-1})\}_{\vec{r}_0}$ into image inputs for the CNN, we use continuous wavelet transforms (CWT) \cite{berthusenLearningCrystalField2021,mallatWaveletTourSignal1999}. These are defined as
\begin{equation}
 W\left(t, h\right) = \frac{1}{\sqrt{s}}\sum_{i=0}^{N_{\omega}-1}\mathcal{D}_{\vec{r}_0}(\omega_{i}) \psi\left(\frac{i-t}{h}\right),\quad \text{with} \quad \psi \left(t\right)= e^{-t^{2}/2}\cos\left(5t\right),
\end{equation}
representing the mother wavelet function in a real Morlet form. Here, the transformation is linearly spaced, with $h=1,2,\dots 65$ being equivalent to the spacing in energy taken in the maps $\mathcal{D}_{\vec{r}_0}(\omega)$. This scale factor is analogous to frequency in Fourier transforms.
Besides $h$, there is a time scale $t$ which is also taken as $t=1,2,\dots 65$, such that $ W\left(t, h\right)$ produces $65\times65$ pixel images. An example of such a ``scaleogram'' is shown in \figref{fig:wavelettransform}.
 
Finally, the LDOS pixel intensities in both maps $\mathcal{D}_{\vec{r}_0}(\omega)$ and $\mathcal{D}_{\omega_0}(\mathbf{r})$ are modified by the addition of Gaussian random noise via $p_{g}\left(z\right)= \exp(-
z^{2}/2\sigma^{2})/\sqrt{2\pi\sigma^{2}} $ with $\sigma=0.31$ \cite{liuSeeingMoirConvolutional2022b}. For $\mathcal{D}_{\vec{r}_0}(\omega)$ images, the noise must be added before the CWT for physical consistency.

\section{Including strain with fixed \texorpdfstring{$\theta_{\epsilon}$}{Lg}, and variations of the ML architecture}\label{app:ml}

\begin{figure} [bt]
	\centering{}\includegraphics[width=\columnwidth]{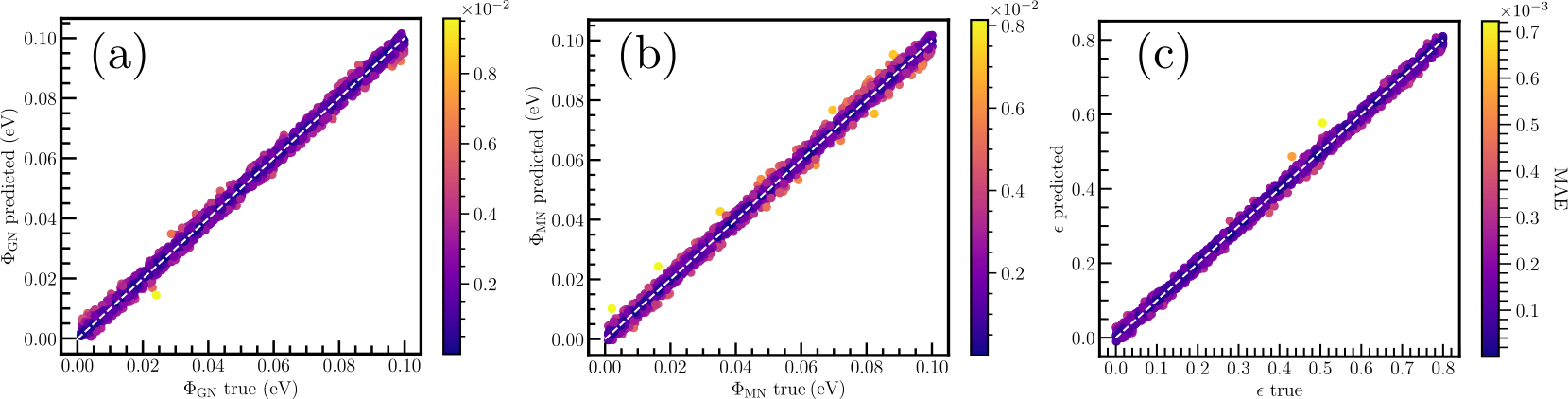}\caption{Predicted versus true values for GN (a), MN (b), and strain intensities (c) for fixed strain angle $\theta_{\epsilon}=0$. Colorbars indicate the MAE for each respective case.}
	\label{fig:strainandnematicappendix}
\end{figure}

In this appendix, we discuss the changes in the performance of the ML procedure when the training data set or the ML architecture are modified.
First, we tested in all cases (Sections \ref{sec:orientation}-\ref{sec:includingstrain}) the performance for the predictions of intensities for fixed angles ($\varphi$, $\alpha$ and $\theta_{\epsilon}$, respectively). For concreteness, we here focus on predicting the microscopic nematic form in the presence of strain (\secref{sec:includingstrain}). The data set for this task is, as before, built by randomly sampling nematic and strain intensities $\Phi^{\text{MN}}, \Phi^{\text{GN}}\in[0.001,0.1]$ eV, and $\epsilon \in [0,0.8]\:\% $. Here, $\theta_{\epsilon}=0$, $\alpha=0$, $\psi_{l}=1$ and $\varphi_{MN}=\varphi_{GN}=\varphi=\pi/3$. In this case, all intensities are easily distinguishable and with high accuracy, see \figref{fig:strainandnematicappendix}. An identical behavior was observed for investigations with fixed $\varphi$ and $\alpha$ for the predictions of the microscopic form of nematicity (\secref{sec:formofnematicity}), showing that this is a general feature of the considered CNN architecture. Naturally, in the absence of outliers, the precision of the CNN can be further increased (i.e., by reaching a lower MAE for the predictions) with increasing size and variability of the data set \cite{cholletDeepLearningPython2021}.

We next address whether the complex architecture in 	\figref{fig:nematicdirector}(a) for each channel is really necessary to solve this inverse problem, or in other words, whether simpler architectures could have the same performance and whether modifications of it could produce significant changes in the predictions. For this, we compared the results from the main text with two other architectures in the case of learning the microscopic form of nematicity (\secref{sec:formofnematicity}): (i) a very simple sequential neural network that takes the images as inputs, followed by a flatten and dense layer which predicts the parameters $\beta=\{\alpha, \Phi_{\text{GN}}, \Phi_{\text{MN}}\}$; (ii) the architecture in \refcite{berthusenLearningCrystalField2021}, which has a similar structure, i.e.,  Conv-Batch-MaxPool channel followed by dense layers, but with different number of filters in each layer; we refer to \refcite{berthusenLearningCrystalField2021} for details of the architecture. We have found that, even if there is some clear correlations between the true and the predicted values of $\alpha$ 
in simple architectures, such as (i), it fails completely on predicting the nematic intensities.  Additionally, (ii) does not lead to any significant improvement in the predictions.

We also investigated the performance of the CNN with respect to hyperparameter optimization for both Sections \ref{sec:formofnematicity} and \ref{sec:includingstrain}. These included using different activation functions (SELU, ELU, LeakyReLU, PReLU, ReLU and Sigmoid) \cite{tensorflow2015-whitepaper}, batch sizes (9, 16, 32, 48 and 96), different number of filters and convolution layers in the Conv-Batch-MaxPool channels, different learning rates ($10^{-1}$, $10^{-2}$, $10^{-3}$ and $10^{-4}$) and optimizers (RMSprop, SGD and ADAM). The architecture described in Sec.~\ref{sec:mlarchitecture} and its variation in \figref{fig:nematicform}(a) already correspond to the optimal configuration. Even though these investigations are not an exhaustive treatment with respect to all possible parameters and correspondent combinations, it shows that certain elements play a major role in the CNN's performance, such as choosing ReLU as activation functions, setting padding to zero in the convolution layers and using a learning rate of $10^{-4}$. Finally, even though the four Conv-Batch-MaxPool channels in the main architecture may not be necessary for simpler cases (e.g. predicting the nematic director with fixed nematic intensities), it is essential for increasing parameters and complexity, such as learning strain and the internal structure of nematicity simultaneously. 

\section{Preprocessing of the experimental data and further implications}\label{app:expdata}

\begin{figure} [bt]
	\centering{}\includegraphics[width=\columnwidth]{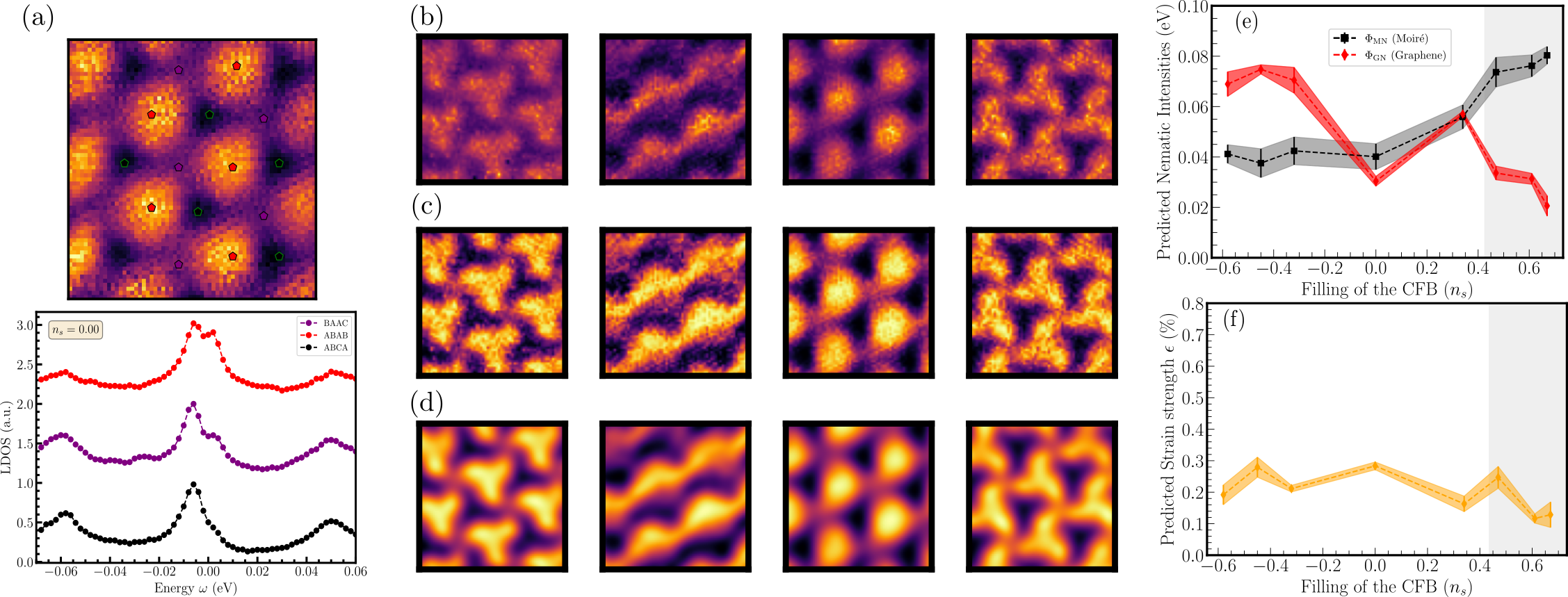}\caption{(a) Location of ABAB, ABCA and BAAC sites in the moiré superlattice for images in $D_{\text{exp}}$. The LDOS channels $\mathcal{D}_{\bm{r}_{0}}(\omega)$ are built by taking an average over the intensities of equivalent sites for each filling $n_{s}$. (b-d) Examples of different preprocessing methods of the $\mathcal{D}_{\omega_0}(\mathbf{r})$ channels: (b) raw data, (c) more contrast and (d) contrast with Gaussian filter. (e-f) Predictions for the same parameters as in \figref{fig:expprediction}(a-b) for different preprocessing procedures; the dots and lines indicate the average values and the shaded region the corresponding standard deviation, see text for more details.}
\label{fig:exp_data_preprocessing}
\end{figure}

The experimental data set $D_{\text{exp}}(\beta_{1}^{\prime},\cdots, \beta_{N_{\text{exp}}}^{\prime})$ consists of $N_{\text{exp}}=8$ samples for fillings of the CFB equal to $n_{s}=\{-0.58, -0.45, -0.32, 0, 0.34, 0.47, 0.61, 0.67\}$. Each sample has
 $\mathcal{D}_{\omega_{0}}(\mathbf{r})$ images in an energy interval of $\omega\in [-100,100]$ meV with resolution of $2$ meV. From these, the  $\mathcal{D}_{\mathbf{r}_{0}}(\omega)$ channels can be calculated by taking an average of intensities at the corresponding BAAC, ABCA and ABAB sites, see \figref{fig:exp_data_preprocessing}(a). 

In order to obtain consistent results, the experimental data set $D_{\text{exp}}$ needs to be fed into the trained CNN as similar as possible to the training data in $D_{\text{th}}$. For this, the preprocessing of $D_{\text{exp}}$ consists of:
\begin{itemize}
\item[(1)] Transforming the experimental  plots $\mathcal{D}_{\mathbf{r}_{0}}(\omega)$ into scaleograms as described in \figref{fig:wavelettransform}. Here, these plots are considered for $\omega \in [-70,60]\,\text{meV}$. This is necessary in order to have scaleograms of $65\times 65$ pixels. In this energy range, these channels contain information about CFB, VFB, RV$_{1}$ and RC$_{1}$. 
\item[(2)] Normalizing each image to have the distribution of pixel intensities with same mean $\mu$ and standard deviation $\sigma$ in both $D_{\text{exp}}$ and $D_{\text{th}}$. Here, we have chosen $\mu=0$ and $\sigma = 1$ \cite{cholletDeepLearningPython2021}. This step is essential to produce meaningful predictions on $D_{\text{exp}}$, since the trained CNN have weights associated to the scale of $D_{\text{th}}$.
\item[(3)] Cropping the images $\mathcal{D}_{\omega_{0}}(\mathbf{r})$ from $D_{\text{exp}}$ such that they show roughly the same number of moiré unit cells as in the corresponding ones in $D_{\text{th}}$. Additionally, the orientations of each of these images in both data setss also need to be consistent pair-wisely, see \figref{fig:exp_data_preprocessing}(b-d) and \figref{fig:expprediction}(c).
\end{itemize}

\begin{figure} [tb]
	\centering{}\includegraphics[width=\columnwidth]{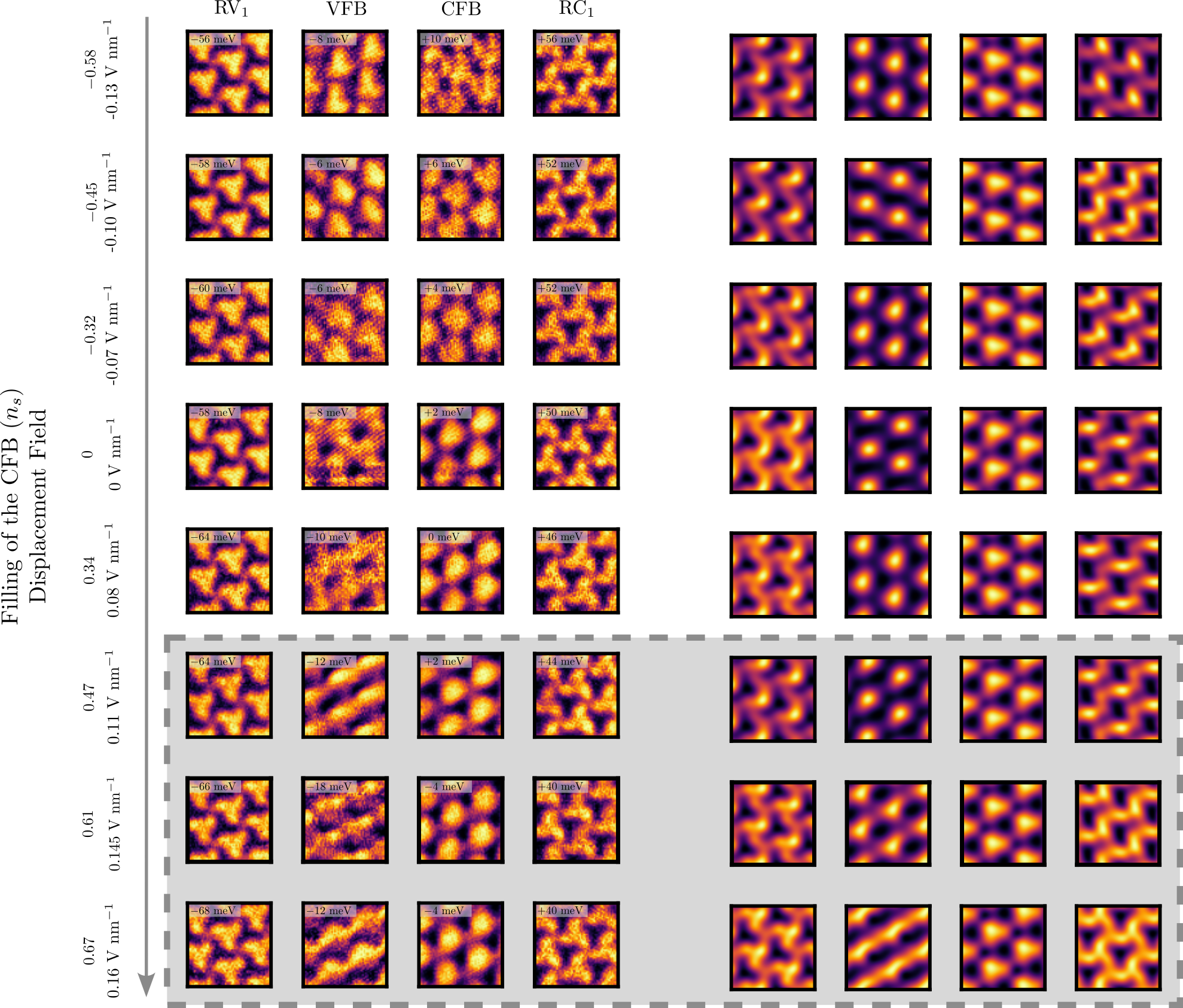}\caption{Comparison between $\mathcal{D}_{\omega_{0}}(\mathbf{r})$ from the experimental data set $D_{\text{exp}}$ (left panel), and the corresponding configuration obtained posteriorly within the continuum model with the predicted $\beta^{\exp}$ from \figref{fig:expprediction}(a-b). For $D_{\text{th}}$ a half-filling fraction of the CFB $(\nu = 0.475)$ corresponds to a chemical potential of $\mu\sim-15$ meV, and the equivalent energies for the RV$_{1}$, VFB, CFB and RC$_{1}$ in the continuum model are found for $\omega_{0}=\{-35,-15,1, 23\}$ meV. These values were chosen for the best possible resemblance of the images $\mathcal{D}_{\omega_{0}}(\mathbf{r})$ in $D_{\text{th}}$ with the ones in $D_{\text{exp}}$, a naturally constrained procedure by the representational power of the theoretical model to the experimental data. The gray box corresponds to the gray regions in \figref{fig:expprediction}(a-b). The experimental images are shown with higher contrast [\figref{fig:exp_data_preprocessing} (c)] for better visual comparison.}
\label{fig:expdata7dospred}
\end{figure}

We have also investigated the influence of additional preprocessing of the $\mathcal{D}_{\omega_{0}}(\mathbf{r})$ channels of the data set $D_{\text{exp}}$ on the predictions. We introduce more contrast to the images [\figref{fig:exp_data_preprocessing}(c)] and reduce noise by smoothing the pixel distribution with a multidimensional Gaussian filter [\figref{fig:exp_data_preprocessing}(d)]. In \figref{fig:exp_data_preprocessing}(e-f), the resulting predictions for the nematicities and strain using these augmented $D_{\text{exp}}$ are shown. Here, every dot represents an average over the predictions on 10 variations of $D_{\text{exp}}$ with Gaussian filter with standard deviations of the Gaussian kernel in $\sigma_{GF}=\{0, 1, 2, 5, 10\}$ with and without higher contrast. The overall behavior described in \secref{sec:experimentaldata} is unaffected by these modifications, but the predictions with the lowest strain intensity in the gray region of \figref{fig:expprediction}(a-b) were found for the raw data inputs [\figref{fig:exp_data_preprocessing}(b)] - see \figref{fig:expdata7dospred}.

We emphasize the importance of the multi-channel CNN architecture in \figref{fig:nematicform}(a) for the predictions in \figref{fig:expprediction}(a-b). While using only $\mathcal{D}_{\omega_{0}}(\mathbf{r})$ at the flat bands already seems to capture the interplay between MN and GN as a function of fillings of the CFB, the addition of channels for the remote bands and scaleograms is crucial to discern the influence of strain and nematicity in the experimental samples. This can be intuitively understood by the relative stability of the remote bands with respect to heterostrain over a wide range of fillings $n_{s}$ in $D_{\text{exp}}$ \cite{rubio-verduMoireNematicPhase2022}. These results indicate that the inclusion of more channels from even more remote bands could potentially produce more accurate predictions for the strain intensity; we leave this for future work.

\section{Applicability of the CNN to different models and moiré systems}\label{app:tbgnematic}
To demonstrate that our ML approach of extracting microscopic parameters based on CNNs with multiple channels works more generally, we here apply it to a different moiré system. To further increase the variability of models studied in this work, we do not use a continuum model but, instead, consider a tight-binding model on the moiré scale that captures the symmetries and topological features of the twisted bilayer graphene (TBG). 

\begin{figure} [bt]
	\centering{}\includegraphics[width=\columnwidth]{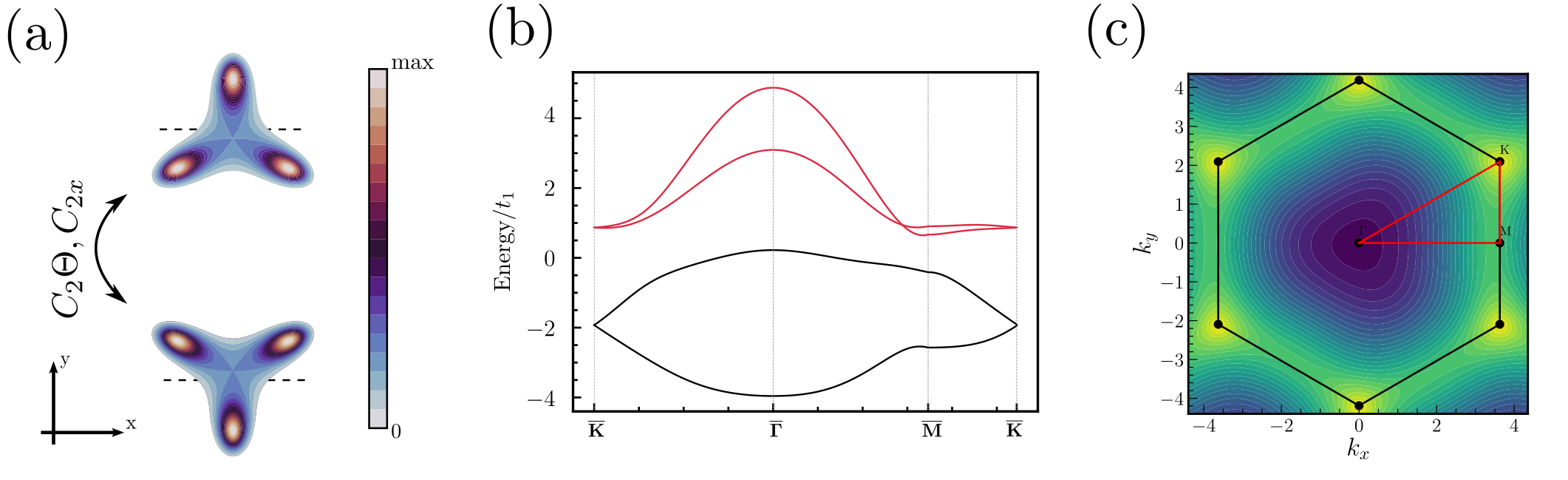}\caption{(a) Illustration of the symmetry properties of the Wannier orbitals defined in \eqref{WannierStates}. In (b), we show the bandstructure of the model along the one-dimensional momentum cut
indicated in red in (c), where the energy of the lowest band is shown as a contour plot. The black
bands in (b) mimic the flat-bands of TBG, while the red lines are just auxiliary bands required
due to the topological obstruction.}
\label{fig:BasicPropertiesModel}
\end{figure}

As is well known \cite{kangSymmetryMaximallyLocalized2018, koshinoMaximallyLocalizedWannier2018, poOriginMottInsulating2018}, the representations of the flat bands of TBG at high-symmetry momenta requires taking a model on the honeycomb lattice. To be able to study the valleys separately, we take the valley quantum number to be conserved such that the associated (fragile) topological obstructions necessitates taking at least four bands \cite{poFaithfulTightbindingModels2019}. We, therefore, place two Wannier orbitals $W_\pm(\vec{r})$ at every site of the honeycomb lattice. We choose them to be invariant under $C_3$ (three-fold rotation perpendicular to the graphene layers) and transform into one-another under $C_{2x}$ (two-fold rotation along $x$) and $\Theta C_2$ (the product of time-reversal and two-fold rotation perpendicular to the layers); this specifies the behavior of the Wannier orbitals under all symmetries of TBG that act within a given valley. Here, our goal is not to provide a quantitatively accurate description of the LDOS of TBG but rather to demonstrate our ML procedure. It is therefore sufficient to take the simple, phenomenological forms of the Wannier states given by
\begin{equation}
    W_{\pm}(\vec{r}) \propto \exp\left(\mp c_{1}y\left(y^{2}-3x^{2}\right)-c_{2}\left(x^{2}+y^{2}\right)^{2}\right), \quad \vec{r} = (x,y)^T, \label{WannierStates}
\end{equation}
which obey the required symmetry constraints, as shown in \figref{fig:BasicPropertiesModel}(a). In \equref{WannierStates}, $c_1$ and $c_2$ are real-valued constants that we set to $\{c_{1},c_{2}\}=\{1.5, 0.7\}$ for concreteness.

Including symmetry-allowed intra-orbital (inter-orbital) hopping processes up to third-nearest (nearest) neighbor leads to a tight-binding model with momentum-space form (for valley $\eta=+$ and a given spin flavor)
\begin{equation}
    \mathcal{H}_{\text{tb}} = \sum_{\vec{k}} c^\dagger_{\vec{k}} h^0_{\vec{k}} c^\pdagger_{\vec{k}}, \quad h^0_{\vec{k}} =
    \begin{pmatrix}
        h^{W_{+}}_{\vec{k}}(\alpha_{2}, \Delta) & h^{C}_{\mathbf{k}} \\
        h^{C\dagger}_{\mathbf{k}} & h^{W_{-}}_{\vec{k}}(-\alpha_{2}, - \Delta) \\
    \end{pmatrix}, \label{TightBindingModel}
\end{equation}
with orbital Hamiltonians
\begin{equation}
    \quad h^W_{\vec{k}}(\alpha_{2}, \Delta) =
    \begin{pmatrix}
        \Delta+ f\left(\alpha_{2},t_{2}\right) &  g\left(t_{1}, t_{3}\right)\\
        g^{\dagger}\left(t_{1}, t_{3}\right) & -\Delta +f\left(-\alpha_{2}, t_{2}\right) \\
    \end{pmatrix}, 
\end{equation}
coupled via 
\begin{equation}
    \quad h^C_{\vec{k}} =
    \begin{pmatrix}
        e^{i\Theta \omega}\omega_{1} &  \omega_{2}\left(1+2e^{i\frac{\sqrt{3}}{2}k_{x}}\cos(k_{y}/2)\right)\\
        \omega_{2}\left(1+2e^{-i\frac{\sqrt{3}}{2}k_{x}}\cos(k_{y}/2)\right) & e^{i\Theta \omega}\omega_{1} \\
    \end{pmatrix}, 
\end{equation}
with 
\begin{equation}
f\left(\alpha_{2}, t_{2}\right) = t_{2}\left(\cos(\mathbf{k}\cdot \mathbf{a}_{2}-\alpha_{2})+\cos(\mathbf{k}\cdot(\mathbf{a}_{1}-\mathbf{a}_{2})-\alpha_{2})+\cos(\mathbf{k}\cdot \mathbf{a}_{1}+\alpha_{2})\right)
\end{equation}
and
\begin{equation}
g\left(t_{1},t_{3}\right) = t_{1}\left(1+e^{i\mathbf{k}\cdot\mathbf{a}_{1}}+e^{i\mathbf{k}\cdot\mathbf{a}_{2}}\right)+t_{3}e^{i\alpha_{3}}\left(e^{-i\mathbf{k}\cdot\left(\mathbf{a}_{2}-\mathbf{a}_{1}\right)}+e^{-i\mathbf{k}\cdot\left(-\mathbf{a}_{2}-\mathbf{a}_{1}\right)}+e^{-i\mathbf{k}\cdot\left(-\mathbf{a}_{2}+\mathbf{a}_{1}\right)}\right),
\end{equation}
where $c^\dagger_{\vec{k}}$ are four-component electronic creation operators with the first two (second two) indices referring to the two sublattices of Wannier orbitals $W_+$ ($W_-$). Here, the two primitive lattice vectors from monolayer graphene are given by $\mathbf{a}_{1}/a=\left(\sqrt{3}, 1\right)/2$ and $\mathbf{a}_{2}/a=\left(\sqrt{3}, -1\right)/2$. In \figref{fig:BasicPropertiesModel}(b,c), we show the band structure for $\{t_{2}/t_{1}, \Delta/t_{1}, \omega_{1}/t_{1}, \omega_{2}/t_{1}, \alpha_{2}, t_{3}/t_{1}, \alpha_{3}, \Theta\omega\}=\{0.6, 0, 0.6, -0.5, 0.3\pi, 0.1, 0.6\pi, 0\}$, which are also used in the ML calculations presented below. We will take the lower two, isolated bands [indicated in black in \figref{fig:BasicPropertiesModel}(b)] as a phenomenological description of the quasi-flat bands of TBG; they exhibit Dirac cones at K and K', which can be shown to have the same chirality---exactly as in TBG.

We next add different forms of nematicity to $h^0_{\vec{k}}$, i.e., $h^0_{\vec{k}} \rightarrow h_{\vec{k}}=h^0_{\vec{k}} + \Delta h_{\vec{k}}$, which have very different structure compared to those discussed in the main text since the model is very different. The nematic order parameter $\bm{\phi}=\left(\phi_{1}, \phi_{2}\right)^{T}=\left(\cos2\varphi, \sin2\varphi\right)^{T}\in \mathbbm{R}^{2}$ will couple as $\Delta h_{\mathbf{k}}=\bm{\phi}\cdot g_{\mathbf{k}}=\phi_{1}g_{1,\mathbf{k}}+\phi_{2}g_{2,\mathbf{k}}$. Here, the matrix-valued functions $g_{l,\mathbf{k}}$ play a similar role as the tensorial form factor $\bm{\phi}_{\sigma,\ell,s, \eta;\:\sigma^{\prime},\ell^{\prime},s^{\prime}, \eta^{\prime}}(\bm{r},\Delta\mathbf{r})$ in the continuum model nematic coupling in \equref{eq:nematorder1}. Denoting by $X_{\vec{k}}$ and $Y_{\vec{k}}$ Brillouin-zone-periodic, real-valued functions that transform as $x$ and $y$ under TBG's point group $D_3$, we can write
\begin{equation}
    \mathbf{g}_{\mathbf{k}} = \alpha_{0}\rho_{0}\sigma_{0} \begin{pmatrix}
        X_{\mathbf{k}} \\ 
        Y_{\mathbf{k}}
    \end{pmatrix} + \alpha_{1}\rho_{0}\sigma_{x} \begin{pmatrix}
        X_{\mathbf{k}} \\ 
        Y_{\mathbf{k}}
    \end{pmatrix}+ \alpha_{2}\rho_{0}\sigma_{y} \begin{pmatrix}
        -Y_{\mathbf{k}} \\ 
        X_{\mathbf{k}}
    \end{pmatrix} + \alpha_{3}\rho_{z}\sigma_{z} \begin{pmatrix}
        -Y_{\mathbf{k}} \\ 
        X_{\mathbf{k}}
    \end{pmatrix},
    \label{eq:minimalmodel2}
\end{equation}
where $\alpha_j \in \mathbbm{R}$ are parameters and $\sigma_j$ ($\rho_j$) are Pauli matrices in Wannier (sublattice) space. Technically, the explicit form of $X_{\vec{k}}$ and $Y_{\vec{k}}$ in each of the four terms in \equref{eq:minimalmodel2} can be different. However, the functional space of possible $X_{\vec{k}}$ and $Y_{\vec{k}}$ is technically infinite dimensional and we will focus only on the leading contribution which then also becomes identical for all four terms in \equref{eq:minimalmodel2} and reads as
\begin{equation}
    (X_{\vec{k}},Y_{\vec{k}}) = \frac{8}{3}\left( \cos k_y - \cos \frac{\sqrt{3}k_x}{2} \cos \frac{k_y}{2} , \sqrt{3} \sin \frac{\sqrt{3}k_x}{2} \sin \frac{k_y}{2}  \right). \label{LowestOrderHarmonic}
\end{equation}
Consequently, there are four parameters, $\beta=\{\alpha_{0}, \alpha_{1}, \alpha_{2}, \alpha_{3}\}$, describing the microscopic form of nematicity in our model. Our goal will be to reconstruct their values from LDOS images, which we compute via 
\begin{equation}
\label{eq:dos_minimalmodel}
    \frac{\text{dI}}{\text{dV}}\left(\mathbf{r},\omega\right)\propto \text{Im}\left[\sum_{j,k, \alpha, \beta}W_{\mathbf{R}_{j\alpha}}\left(\mathbf{r}\right) G^{R}_{\alpha \beta}\left(\mathbf{R}_{j}-\mathbf{R}_{k}, \omega\right)W_{\mathbf{R}_{k\beta}}^{*}\left(\mathbf{r}\right) \right] = \mathcal{D}\left(\mathbf{r},\omega \right)
\end{equation}
with 
\begin{equation}
    G^{R}_{\alpha \beta}=\left(\mathbf{R}-\mathbf{R}^{\prime},\omega\right) = \frac{1}{V}\sum_{\mathbf{k}}e^{i\mathbf{k}\left(\mathbf{R}-\mathbf{R}^{\prime}\right)}\lim_{\eta\rightarrow 0^{+}}\left(\frac{1}{\omega-h_{\mathbf{k}}+i\eta}\right)_{\alpha,\beta}.
\end{equation}
The indices $\alpha$ and $\beta$ of $W_{\mathbf{R}_{j\alpha}}$ in \equref{eq:dos_minimalmodel}
correspond to four different realizations of the Wannier functions in each unit cell $\vec{R}_j$, as each of the two
orbitals $W_{\pm}$ can be placed on each of the two sublattices.

\begin{figure} [tb]
\centering{}\includegraphics[width=\columnwidth]{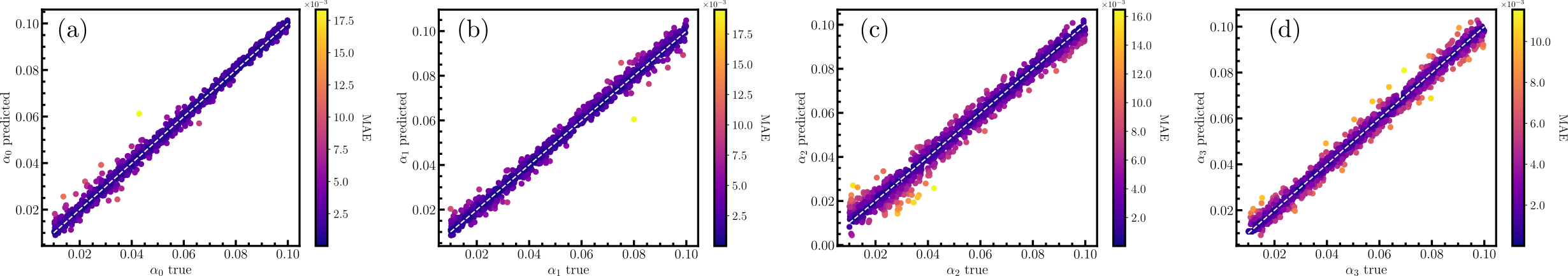}\caption{Predicted versus true values for the nematic paramters $\alpha_{0}$ (a), $\alpha_{1}$ (b), $\alpha_{2}$ (c) and $\alpha_{3}$ (d) defined in \eqref{eq:minimalmodel2} for the minimal model in \equref{TightBindingModel}. As before, colorbars indicate the MAE for each respective case.}
\label{fig:minimalmodelpred}
\end{figure}

To reconstruct $\beta=\{\alpha_{0}, \alpha_{1}, \alpha_{2}, \alpha_{3}\}$, we consider a variation of the ML architecture in \figref{fig:nematicform}(a) with four channels for $\mathcal{D}_{\omega_{0}}(\mathbf{r})$ with $\omega_{0}/t_{1}=\{-2, -1, 1, 2\}$, and one for the scaleograms from $\mathcal{D}_{\mathbf{r}_{0}}(\omega)$.
 The complete data set consists of 12000 images
which are divided into training (78.5\%), validation (15\%)
and test (6.5\%) subgroups. These are generated with randomly sampled $\alpha_{j} \in [0.01, 0.1]$ for $j=0,\cdots, 3$ and for a fixed nematic director $\varphi = 5\pi/6$. All the images in the data set were modified by the addition of Gaussian noise with a standard deviation of $\sigma=0.05$. In \figref{fig:minimalmodelpred}(a-d), one can see that all four parameters can be accurately predicted. Additionally, we have also observed that even training the CNN with only a single $\mathcal{D}_{\mathbf{r}_{0}}\left(\omega\right)$ channel in this case is sufficient to also yield very good predictions, evidencing the fundamental role of point spectra as an additional source of information. These results indicate that the framework proposed in this work could be successfully applied to a plethora of correlated phenomena and different moiré systems.

\end{appendix}
\end{document}